\documentclass[aps,pra,twocolumn,reprint,superscriptaddress]{revtex4-2}
\usepackage{graphicx}
\usepackage[english]{babel}
\usepackage[inline]{enumitem}
\usepackage{amsmath, amssymb, amsfonts}
\usepackage[caption=false,labelformat=simple]{subfig}% For subfigure
\usepackage{bm}
\usepackage{xcolor}
\usepackage[colorlinks=true,linkcolor=blue]{hyperref}
\usepackage[normalem]{ulem}

\newlength{\figwidth}
\newlength{\lfig}
\newlength{\sfig}
\begin{document}
\title{Shielding collisions of ultracold CaF molecules with static electric fields}

\author{Bijit Mukherjee}
\affiliation{Joint Quantum Centre (JQC) Durham-Newcastle, Department of Chemistry, Durham University, South Road, Durham, DH1 3LE, United Kingdom.}
\author{Matthew D. Frye}
\affiliation{Joint Quantum Centre (JQC) Durham-Newcastle, Department of Chemistry, Durham University, South Road, Durham, DH1 3LE, United Kingdom.}
\author{C. Ruth Le Sueur}
\affiliation{Joint Quantum Centre (JQC) Durham-Newcastle, Department of Chemistry, Durham University, South Road, Durham, DH1 3LE, United Kingdom.}
\author{Michael R. Tarbutt}
\address{Centre for Cold Matter, Blackett Laboratory, Imperial College London, Prince Consort Road, London SW7 2AZ UK}
\author{Jeremy M. Hutson}
\email{j.m.hutson@durham.ac.uk} \affiliation{Joint Quantum Centre (JQC)
Durham-Newcastle, Department of Chemistry, Durham University, South Road,
Durham, DH1 3LE, United Kingdom.}

\date{\today}

\begin{abstract}
We study collisions of ultracold CaF molecules in strong static electric fields. These fields allow the creation of long-range barriers in the interaction potential, effectively preventing the molecules from reaching the short-range region where inelastic and other loss processes are likely to occur. We carry out coupled-channel calculations of rate coefficients for elastic scattering and loss. We develop an efficient procedure for including energetically well-separated rotor functions in the basis set via a Van Vleck transformation. We show that shielding is particularly efficient for CaF and allows the rate of two-body loss processes to be reduced by a factor of $10^7$ or more at a field of 23 kV/cm. The loss rates remain low over a substantial range of fields. Electron and nuclear spins cause strong additional loss in some small ranges of field, but have little effect elsewhere. These results pave the way for evaporative cooling of CaF towards quantum degeneracy.
\end{abstract}

\maketitle

\section{Introduction}

Ultracold molecules have many potential applications, ranging from quantum simulation \cite{Gorshkov:2011, Baranov:2012} and quantum computing \cite{DeMille:2002, Yelin:2006} to the creation of novel quantum phases \cite{Buechler:2007, Lechner:2013}. There is particular interest in polar molecules, which can have long-range anisotropic interactions resulting from their permanent dipoles. A variety of such molecules have been produced at microkelvin temperatures by association of pairs of atoms \cite{Ni:KRb:2008, Takekoshi:RbCs:2014, Molony:RbCs:2014, Park:NaK:2015, Guo:NaRb:2016, Rvachov:2017}, or by direct laser cooling \cite{Truppe:MOT:2017, McCarron:2018, Anderegg:2018, Cheuk:2018, Caldwell:2019, Ding:2020}.

Many applications of ultracold molecules need high phase-space densities. For atoms, this is usually achieved by evaporative or sympathetic cooling \cite{Ketterle:1996, Myatt:1997}. However, high-density samples of ultracold molecules usually undergo collisional loss, due to a variety of short-range mechanisms that may include two-body inelastic or reactive collisions \cite{Ospelkaus:react:2010}, three-body collisions \cite{Mayle:2013} or laser-induced loss \cite{Christianen:laser:2019}. There is therefore much interest in \emph{shielding} collisions of ultracold molecules to prevent colliding pairs reaching short range. This can be achieved by confining molecules to two dimensions, with an electric field perpendicular to the plane, so that repulsive dipole-dipole interactions dominate \cite{Micheli:2007, Quemener:2011, deMiranda:2011}. Alternatively, in three dimensions, it requires \emph{engineering} repulsive interactions based on the dipole-dipole interaction.

There have been theoretical proposals to achieve 3D shielding using static electric fields \cite{Avdeenkov:2006, Wang:dipolar:2015, Quemener:2016, Gonzalez-Martinez:adim:2017}, near-resonant microwaves \cite{Karman:shielding:2018, Karman:shielding-imp:2019}, or lasers \cite{Xie:optical:2020}. Both microwave shielding and static-field shielding have been demonstrated experimentally \cite{Matsuda:2020, Li:KRb-shield-3D:2021, Anderegg:2021, Schindewolf:NaK-degen:2022}.
In this paper, we focus on static-field shielding of bosonic CaF, although our results also apply to similar molecules. CaF is of interest because it has been laser-cooled to a few $\mu$K and confined in magnetic and optical traps \cite{Anderegg:2018, Cheuk:2018, Caldwell:2019}. In the optical traps, phase-space densities are reaching the regime where evaporative cooling becomes feasible, especially when the molecules are electrically polarized so that elastic collision rates are enhanced by dipolar interactions \cite{Bohn:BCT:2009}. However, it is known from studies using tweezer traps that ground-state CaF molecules undergo two-body collisional loss with a loss rate constant of $(7 \pm 4) \times 10^{-11}$ cm$^{3}$ s$^{-1}$ \cite{Cheuk:2020}. This may be due to the reaction 2CaF $\rightarrow$ CaF$_2$ + Ca, which is exothermic \cite{Vasiliu:2010} and barrierless \cite{Sardar:2022}, or to the formation and subsequent loss of complexes. Evaporative cooling is impossible in the presence of such fast, destructive collisions.

It is thus important and timely to consider the effectiveness of electric shielding for CaF and similar systems. Here, we present coupled-channel calculations to demonstrate that shielding with a static electric field is likely to be extremely effective for CaF. We calculate cross sections for elastic scattering and trap loss as a function of electric field and collision energy. We show how s-wave and higher partial waves contribute in both cases. We then use these results to evaluate the likely effectiveness of evaporative cooling in this system.

The structure of the paper is as follows. Section \ref{sec:theory} describes our coupled-channel treatment, together with a discussion of basis-set size and an efficient way of extending the basis set using a Van Vleck transformation. Section \ref{sec:results} describes our results for elastic scattering and loss processes, first in the spin-free case and then including the effects of electron and nuclear spins and of magnetic field. Section \ref{sec:conc} presents our conclusions. Finally, the Appendix presents a detailed discussion of the convergence of the calculations, together with an analysis of resonance oscillations that can occur in some cases.

\section{Theory}
\label{sec:theory}

\subsection{Coupled-channel approach for the spin-free case}

\begin{figure}[tbp]
\begin{center}
\includegraphics[width=\figwidth,clip]{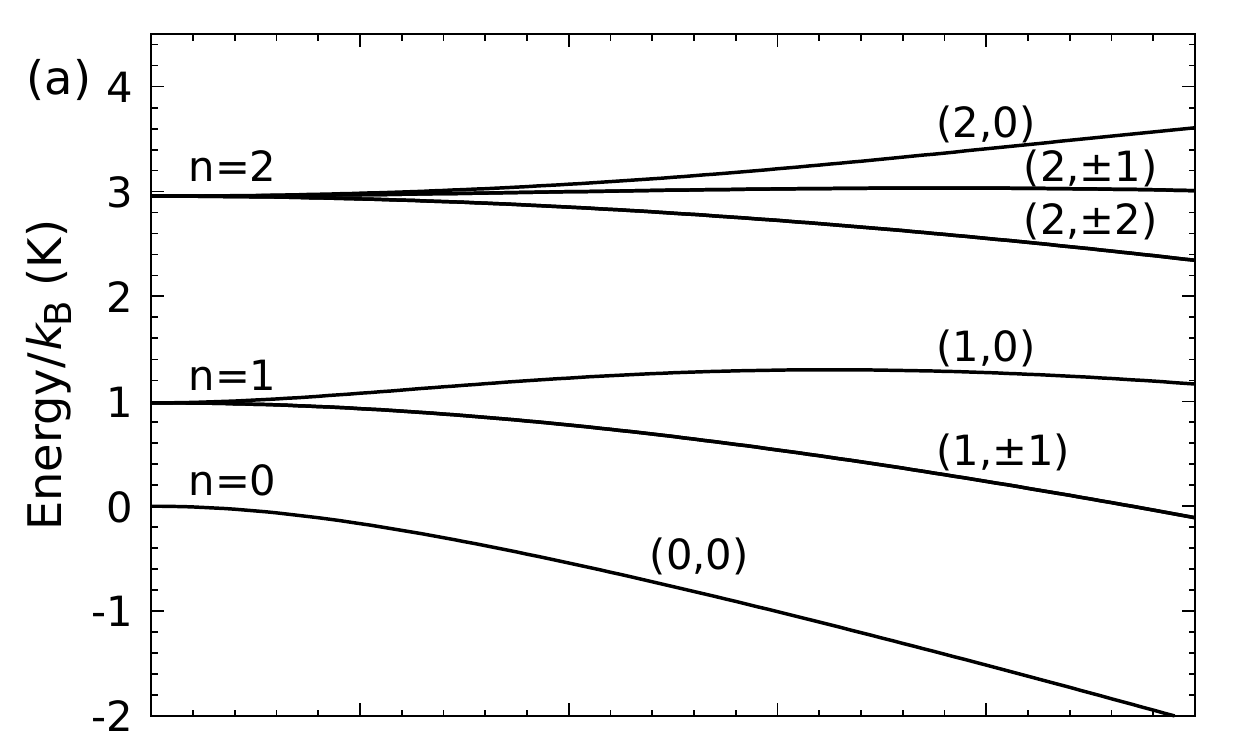}
\includegraphics[width=\figwidth,clip]{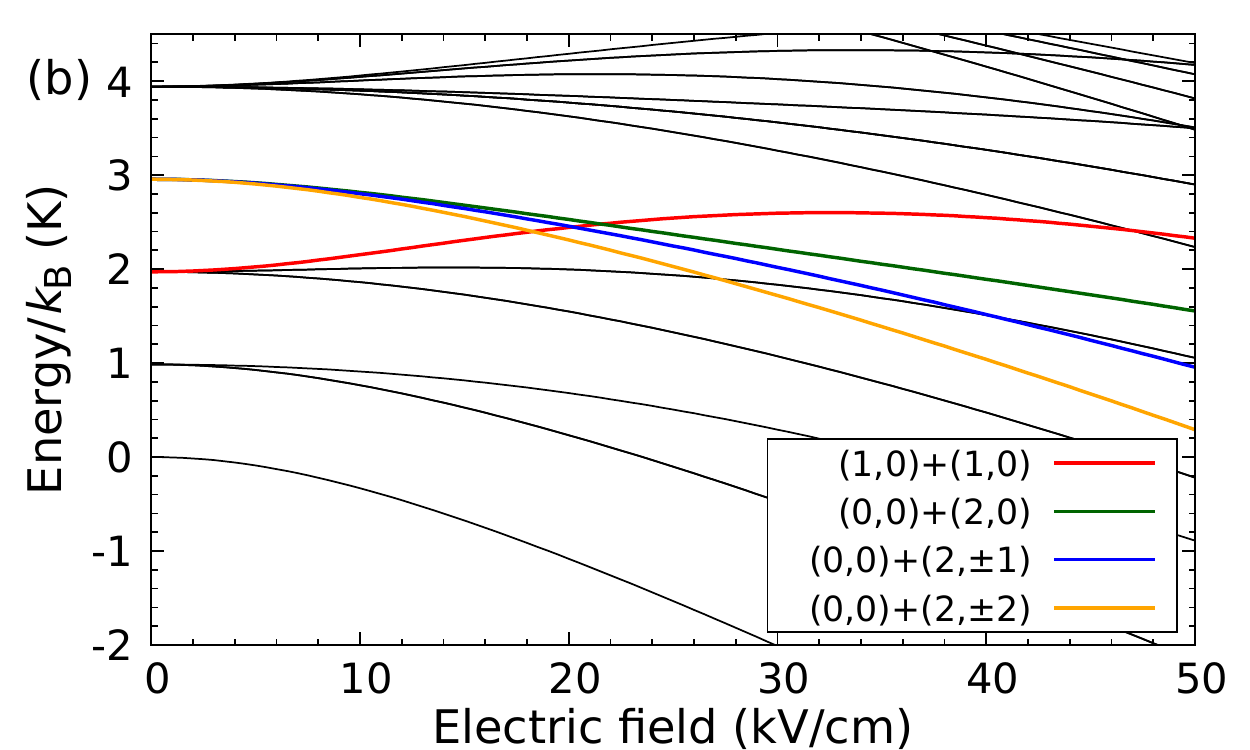}
\caption{Energy of (a) a single CaF molecule and (b) a pair of CaF molecules as a function of electric field, neglecting electron and nuclear spin.
The initial pair state ($\tilde{n},m_n)$ = (1,0)+(1,0) is shown in red. The states (0,0)+(2,0), (0,0)+(2,$\pm1$) and (0,0)+(2,$\pm2$) are shown in green, blue, and orange, respectively, and cross the initial state between 18 and 22 kV/cm.}
\label{fig:CaF_Stark}
\end{center}
\end{figure}

We begin by treating each CaF molecule as a rigid rotor with a dipole moment $\boldsymbol{\mu}_k$ oriented along its internuclear axis. Electron and nuclear spins are initially neglected, but will be considered later. The Hamiltonian for a single molecule $k$ is
\begin{equation}
\hat{h}_k = b_k\hat{\boldsymbol{n}}_k^2 - \boldsymbol{\mu}_k \cdot \boldsymbol{F},
\label{eq:ham-Stark}
\end{equation}
where $\hat{\boldsymbol{n}}_k$ is the operator for molecular rotation, $b_k$ is the rotational constant, and $\boldsymbol{F}$ is an applied electric field along the $z$ axis. For $^{40}$Ca$^{19}$F, $b/h \approx 10.267$\,GHz and $|\boldsymbol{\mu}|=3.07$\,D. Figure~\ref{fig:CaF_Stark}(a) shows the single-molecule energy levels as a function of electric field; we label the levels $(\tilde{n},m_n)$; here $\tilde{n}$ is a quantum number that correlates with the free-rotor quantum number $n$ at zero field and $m_n$ represents the conserved projection of $n$ onto the $z$ axis. Figure~\ref{fig:CaF_Stark}(b) shows the corresponding energy of a \emph{pair} of noninteracting CaF molecules.

The dipole-dipole interaction between two molecules takes the form
\begin{equation}
\hat{H}_\textrm{dd} = [3(\boldsymbol{\mu}_1\cdot\hat{\boldsymbol{R}}) (\boldsymbol{\mu}_2\cdot\hat{\boldsymbol{R}}) - \boldsymbol{\mu}_1\cdot\boldsymbol{\mu}_2] / (4\pi\epsilon_0 R^3),
\end{equation}
where $R$ is the intermolecular distance and $\hat{\boldsymbol{R}}$ is a unit vector along the intermolecular axis. Shielding may occur when two pair states that are connected by $\hat{H}_\textrm{dd}$ are close enough in energy that they are strongly mixed. Two molecules that collide on the upper curve then experience a repulsive potential curve proportional to $1/R^3$. In Fig.~\ref{fig:CaF_Stark}(b), this can occur when two molecules in the state (1,0) collide at fields just above 21.55 kV/cm, where (1,0)+(1,0) lies just above (0,0)+(2,0). It can also occur just above 20.20 kV/cm, where (1,0)+(1,0) lies just above (0,0)+(2,$\pm1$). In the remainder of this paper, we focus on collisions of pairs of molecules in the state (1,0).

The Hamiltonian for a colliding pair of molecules is
\begin{equation}
\hat{H} = \frac{\hbar^2}{2\mu}\left( -R^{-1} \frac{d^2}{dR^2} R + \frac{\hat{\boldsymbol{L}}^2}{R^2} \right)
+ \hat{h}_1 + \hat{h}_2  + V_\textrm{int}, 
\label{eq:ham-pair}
\end{equation}
where $\mu$ is the reduced mass, $\hat{\boldsymbol{L}}$ is the operator for relative rotation and $V_\textrm{int}$ is the interaction potential. The total wavefunction is expanded
\begin{equation} \Psi(R,\hat{\boldsymbol{R}},\hat{\boldsymbol{r}}_1,\hat{\boldsymbol{r}}_2)
=R^{-1}\sum_j\Phi_j(\hat{\boldsymbol{R}},\hat{\boldsymbol{r}}_1,\hat{\boldsymbol{r}}_2)\psi_{j}(R), \label{eq:expand}
\end{equation}
where $\hat{\boldsymbol{r}}_k$ is a unit vector along the axis of molecule $k$.
We use a basis set of functions $\{\Phi_j\}$,
\begin{equation}
\Phi_j = \phi^{\tilde{n}_1}_{m_{n1}}(\hat{\boldsymbol{r}}_1) \phi^{\tilde{n}_2}_{m_{n2}}(\hat{\boldsymbol{r}}_2) Y_{LM_L}(\hat{\boldsymbol{R}}),
\label{eq:basis-spin-free}
\end{equation}
symmetrized for exchange of identical bosons. Here $\phi^{\tilde{n}_1}_{m_{n1}}(\hat{\boldsymbol{r}}_1)$ and $\phi^{\tilde{n}_2}_{m_{n2}}(\hat{\boldsymbol{r}}_2)$ are field-dressed rotor functions that diagonalize $\hat{h}_1$ and $\hat{h}_2$, respectively, and $Y_{LM_L}(\hat{\boldsymbol{R}})$ are spherical harmonics that are the eigenfunctions of $\hat{\boldsymbol{L}}^2$. This produces a set of coupled differential equations, which are solved as described below.

The field-dressed functions $\phi^{\tilde{n}}_{m_{n}}(\hat{\boldsymbol{r}})$ are themselves expanded in free-rotor functions $Y_{nm_n}(\hat{\boldsymbol{r}})$. An advantage of using field-dressed basis functions in the expansion (\ref{eq:expand}) is that it is possible to use a large value of $n_\textrm{max}$ in solving Eq.\ (\ref{eq:ham-Stark}) and then a smaller value of $\tilde{n}_\textrm{max}$ for the basis set (\ref{eq:basis-spin-free}) used to solve the coupled equations. The computer time taken to solve the coupled equations is determined by $\tilde{n}_\textrm{max}$, and values of $n_\textrm{max}>\tilde{n}_\textrm{max}$ result in a negligible increase in total computer time.

The projection of the total angular momentum, $M_\textrm{tot}=m_{n1}+m_{n2}+M_L$, is a conserved quantity. We therefore solve the coupled equations separately for each value of $M_\textrm{tot}$.

\subsection{Interaction potential}

The full interaction potential between two CaF molecules is very deep and strongly anisotropic at short range \cite{Sardar:2022}. However, shielding occurs due to dipole-dipole interactions that occur at intermolecular distances $R\gg 100\ a_0$. At these distances, the chemical interactions that dominate at short range make very little contribution, and they are neglected in the present paper. However, there are significant effects due to dispersion interactions, which are proportional to $R^{-6}$ at long range. These are of two types. Rotational dispersion interactions arise from matrix elements of $\hat{H}_\textrm{dd}$ off-diagonal in monomer rotational quantum numbers, which are included directly in the coupled equations. In addition, there are electronic dispersion interactions, arising from dipole-dipole matrix elements off-diagonal in electronic state. We take these into account through an additional interaction $V_\textrm{disp}^\textrm{elec} = -C_6^\textrm{elec}/R^6$, with $C_6^\textrm{elec} \approx 2300\ E_\textrm{h} a_0^6$.

\subsection{Van Vleck transformation and adiabatic curves}
\label{sec:VV}

The interaction potential $V_\textrm{int}$ is dominated at long range by $\hat{H}_\textrm{dd}$, with shorter-range contributions from higher-order multipolar interactions, dispersion forces and chemical bonding interactions. $\hat{H}_\textrm{dd}$ causes strong mixing of partial waves $L$, even at quite long range. Furthermore, incoming partial waves with $L \gg 0$ make substantial contributions to elastic cross sections for dipolar scattering, even for very low collision energies \cite{Bohn:BCT:2009}. Because of this, basis sets with large values of $L_\textrm{max}$ are needed. The details of the convergence are described in the Appendix, but for illustration we use basis sets with $L$ up to $L_\textrm{max}=20$ in this section.

In the presence of an electric field, total parity is not conserved. The only quantities that are fully conserved are the exchange symmetry for identical particles (which must be $+1$ for $^{40}$Ca$^{19}$F, which is a composite boson) and $M_\textrm{tot}$. However, $(-1)^L$ is also conserved if the only terms in $V_\textrm{int}$ are $V_\textrm{dd}$ and $V_\textrm{disp}^\textrm{elec}$. Even with the latter restriction, the resulting basis sets are very large; for example, for $\tilde{n}_\textrm{max}=5$ and $L_\textrm{max}=20$, there are $N=6240$ channels with $M_\textrm{tot}=0$ and even $L$ in the expansion (\ref{eq:expand}). Coupled-channel calculations take computer time approximately proportional to $N^3$, and are very challenging for such large basis sets.

To circumvent this issue, we solve coupled equations that include explicitly only a small number of pair functions $(\tilde{n}_1,m_{n1})$+$(\tilde{n}_2,m_{n2})$. The remaining pair functions are included through an effective Hamiltonian that takes account of $H_\textrm{dd}$ through a Van Vleck transformation \cite{VanVleck:1928, Kemble:1937}. The full set of channels is partitioned into two classes, denoted class 1 (labels a, b, $\ldots$) and class 2 (labels $\alpha$, $\beta$, $\ldots$), such that no channel in class
2 is asymptotically close in energy to any channel in class 1. The channels in class 1 are included explicitly in the coupled-channel calculations, while those in class 2 are included perturbatively. Formally, we perform a unitary transformation such that the matrix elements of $\hat{H}_\textrm{dd}$ between channels in class 1 and class 2 vanish up to first order. In second-order perturbation theory, this contributes matrix elements \emph{between} the channels in class 1 of the form
\begin{align}
\langle a &| \hat{H}_\textrm{dd,VV} | b \rangle \nonumber\\
&= \sum_\alpha \frac{1}{2} \left[
\frac{\langle a | \hat{H}_\textrm{dd} | \alpha \rangle
\langle \alpha | \hat{H}_\textrm{dd} | b \rangle} {(E_a-E_\alpha)}
+ \frac{\langle a | \hat{H}_\textrm{dd} | \alpha \rangle
\langle \alpha | \hat{H}_\textrm{dd} | b \rangle} {(E_b-E_\alpha)}
\right].
\end{align}
We make the further approximation of replacing the energies in the denominators with their asymptotic values, so that they are independent of $R$.
Since $\hat{H}_\textrm{dd}$ is proportional to $R^{-3}$, $\hat{H}_\textrm{dd,VV}$ is proportional to $R^{-6}$. The selection rules for matrix elements of $\hat{H}_\textrm{dd}$ are $\Delta L=0,\pm2$ and $\Delta M_L=0,\pm1,\pm2$, so that those of $\hat{H}_\textrm{dd,VV}$ are $\Delta L=0,\pm2,\pm4$ and $\Delta M_L=0,\pm1,\pm2,\pm3,\pm4$.

\begin{figure}[tbp]
\includegraphics[width=0.49\textwidth]{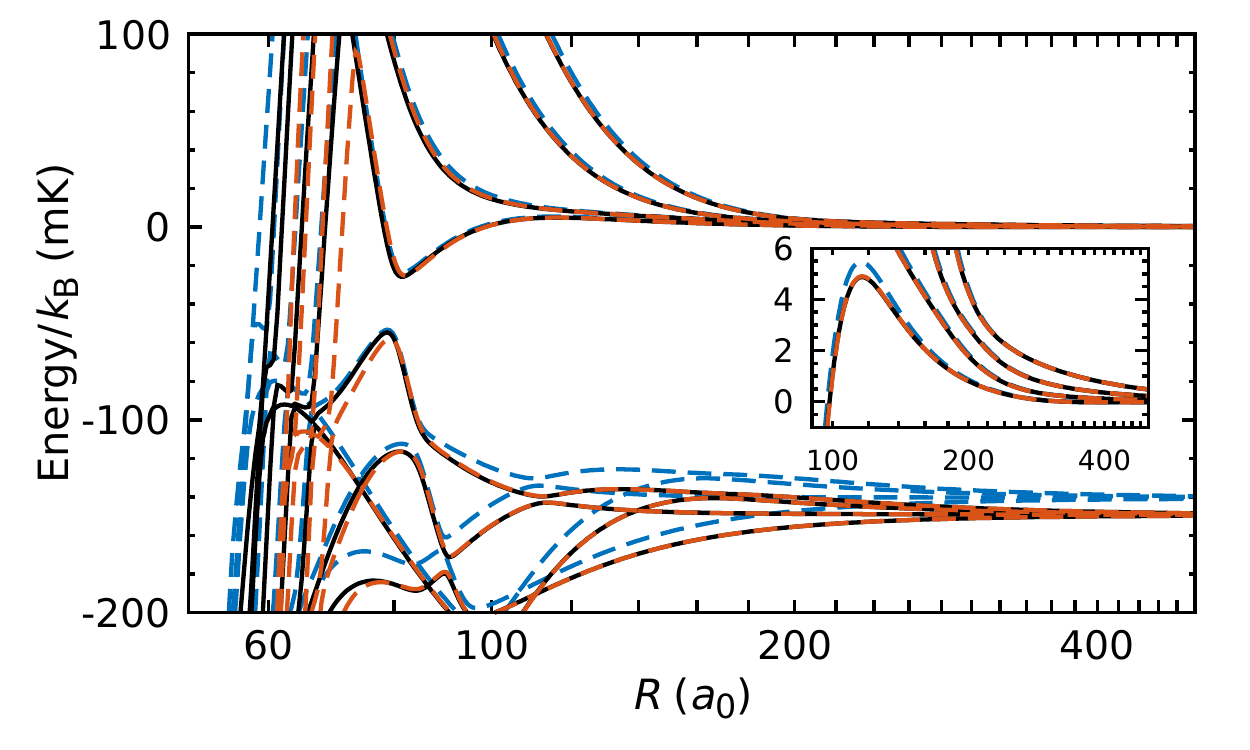}
\caption{Adiabats correlating with pair levels (1,0)+(1,0) and (0,0)+(2,0), calculated for an electric field of 24.5 kV/cm with $L_\textrm{max}=6$ by full diagonalization with $n_\textrm{max}=5$ (black solid lines), $n_\textrm{max}=3$ (blue dashed lines) and with a Van Vleck transformation including only pair levels up to $\tilde{n}=2$ in class 1, but with all the additional levels up to $n_\textrm{max}=5$ in class 2 (dashed orange lines). The $R$-axis is logarithmic, with tick marks separated by 20~$a_0$. The inset shows an expanded view of the adiabats correlating with $L=0$, 2, 4, 6 at the threshold (1,0)+(1,0) near the long-range barrier for incoming $L=0$.
}
\label{fig:adiabats}
\end{figure}

Shielding may be understood qualitatively in terms of effective potential curves obtained by diagonalizing $\hat{h}_1+\hat{h}_2+V_\textrm{int}$ at fixed values of $R$. To a first approximation, collisions occur on these ``adiabats", although there are also transitions between them that are fully accounted for in coupled-channel calculations. Figure \ref{fig:adiabats} shows the adiabats correlating with (1,0)+(1,0) and (0,0)+(2,0), for an electric field of 24.5 kV/cm, where shielding is moderately effective, calculated in several ways as described below. When (0,0)+(2,0) lies slightly below (1,0)+(1,0), as at this field, the adiabats for (1,0)+(1,0) are repulsive at distances of a few hundred bohr due to mixing with the lower threshold; it is this repulsion that can prevent molecules reaching short range and produce shielding. A particularly important feature in Fig.\ \ref{fig:adiabats} is the barrier in the incoming channel with $L=0$, with finite height and width as shown in the inset.

The black solid lines in Fig.~\ref{fig:adiabats} show adiabats calculated by direct diagonalization using a basis set with $\tilde{n}_\textrm{max}=n_\textrm{max}=5$ and $L_\textrm{max}=6$. The blue dashed lines show adiabats calculated with a smaller basis set with $\tilde{n}_\textrm{max}=n_\textrm{max}=3$; it may be seen that this does not accurately reproduce the barrier in the incoming channel with $L=0$, and also gives slightly incorrect threshold energies. The dashed orange lines show adiabats calculated with a Van Vleck transformation, with only levels up to $\tilde{n}=2$ in class 1, but with all the additional levels up to $\tilde{n}_\textrm{max}=5$ included in class 2. It may be seen that the Van Vleck transformation faithfully represents the full adiabats for $R>80\ a_0$, including the height and width of the barrier. There are some differences at shorter range, mainly due to channels that come down from the higher thresholds, but these do not make important contributions to shielding.

Use of a Van Vleck transformation allows an enormous reduction in the number of channels needed in coupled-channel calculations. The spin-free calculations described below use a basis set with functions up to $\tilde{n}=2$ in class 1. With $L_\textrm{max}=20$ this requires 455 basis functions in the coupled-channel calculations, compared to 6240 needed for the full basis set with $\tilde{n}_\textrm{max}=5$. Furthermore, Van Vleck transformations that include only a very few pair levels in class 1 still provide qualitatively accurate results, as described in the Appendix.

When a Van Vleck transformation is used, the computer time taken to solve the coupled equations is determined almost entirely by the number of functions in class 1. There is then no further advantage in choosing $\tilde{n}_\textrm{max}<n_\textrm{max}$, as there is without a Van Vleck transformation. In the remainder of this paper, therefore, we use basis sets with $\tilde{n}_\textrm{max}=n_\textrm{max}$, but with only a subset of the resulting functions included in class 1 and thus in the coupled equations.

\subsection{Trap loss}

Colliding molecules may be lost from a trap in two ways. First, colliding pairs may undergo a transition to a lower-lying pair state. In this case both molecules acquire kinetic energy that is almost always larger than the trap depth, and are lost from the trap. We refer to this as inelastic loss. Secondly, any pairs that penetrate through the engineered repulsive barrier and reach small intermolecular distance are also likely to be lost. This may occur by a variety of mechanisms, including short-range inelasticity, laser absorption, or three-body collisions. We refer to this as short-range loss and to the sum of inelastic and short-range loss as total loss.

To model these processes, we solve the coupled equations with a fully absorbing boundary condition at short range, as in Refs.\ \cite{Clary:1987, Janssen:PhD:2012}. We use log-derivative propagators \cite{Manolopoulos:1986, Alexander:1987} adapted to co-propagate two linearly independent solutions for each channel, and use these to construct traveling-wave solutions with no outgoing part at a distance $R_\textrm{absorb}$. This produces a non-unitary S matrix that is used to produce separate cross sections $\sigma_\textrm{el}$, $\sigma_\textrm{inel}$ and $\sigma_\textrm{short}$ for elastic scattering, inelastic scattering and short-range loss, respectively. The expressions for the cross sections in terms of S-matrix elements are given in the Appendix. The corresponding rate coefficients $k$ at collision energy $E_\textrm{coll}$ are related to the cross sections $\sigma$ through $k = v \sigma$, where $v=(2E_\textrm{coll}/\mu)^{1/2}$. Thermally averaged rate coefficients require further averaging over a Maxwell-Boltzmann distribution, but that is not performed here.

At the electric fields of interest for shielding, dipole-dipole and isotropic dispersion interactions generally dominate the collision physics at distances greater than $100\ a_0$. At shorter distances, however, other forces that are not included here start to contribute. These include dipole-quadrupole interactions, the anisotropy of electronic dispersion forces and (at yet shorter distances) chemical bonding. Nevertheless, under some circumstances there are interesting resonance effects due to states confined between 60 and $100\ a_0$, as described in the Appendix. To capture these effects, we place the fully absorbing boundary at $R_\textrm{absorb}=50\ a_0$, effectively assuming that all collisions that reach that distance produce trap loss. This is a reasonably conservative approximation, although it does not give a strict upper bound to loss, because reflections at shorter range can in principle cause enhanced loss through interference effects.

\subsection{Electron and nuclear spins}

All experiments on laser-cooled CaF so far have been carried out on $^{40}$Ca$^{19}$F, where $^{19}$F has nuclear spin $i=1/2$ and $^{40}$Ca has no nuclear spin. In addition, there is an electron spin $s=1/2$. The spins contribute several extra terms in the molecular Hamiltonian due to fine and hyperfine interactions. To account for them in our coupled-channel calculations, we supplement the field-dressed spin-free functions $|\tilde{n},m_n\rangle$ with functions for the electron and nuclear spins. This will be described in Sec. \ref{sec:spins} below.

\section{Results}
\label{sec:results}

\subsection{The spin-free case}

This section will explore CaF+CaF collisions as a function of electric field, neglecting electron and nuclear spins.
Initially we present rate coefficients at $E_\textrm{coll}/k_\textrm{B}=10\,\textrm{nK}$ and $10\,\mu\textrm{K}$. The latter is slightly above the lowest temperature of 5\,$\mu$K so far achieved for CaF \cite{Cheuk:2018, Caldwell:2019}, so is a likely starting point for evaporative cooling. The former is close to the regime of limitingly low energy. In this regime, the cross sections for inelastic scattering and short-range loss are proportional to $1/v$, so the corresponding rate coefficients are independent of energy. The elastic cross sections, however, are independent of energy in the low-energy limit, so $k_\textrm{el} \propto E_\textrm{coll}^{1/2}$. Further details of the energy dependence and its origins are given in Sec. \ref{sec:Ecoll}.

Before presenting results, we consider the convergence of the calculations with respect to basis set. All calculations use a rotor basis set with $\tilde{n}_\textrm{max}=n_\textrm{max}=5$, which is very well converged. However, only a subset of the field-dressed pair functions are included in the class-1 basis set; the remainder are accounted for by a Van Vleck transformation. As shown in the Appendix, even very small class-1 basis sets give qualitatively correct results across the whole range of fields of interest. However, small rotor basis sets can introduce oscillations due to resonance effects. These oscillations are suppressed with larger class-1 basis sets. Except where otherwise stated, we use a basis set with all combinations of field-dressed rotor functions up to $\tilde{n}=2$ included in class 1. This is referred to below as the ``large" rotor basis set.

As shown in the Appendix, the elastic cross sections converge quite fast with respect to $L_\textrm{max}$, but the loss cross sections converge much more slowly. In the remainder of this section, we use basis sets with $L_\textrm{max}=20$, chosen to give convergence of both elastic and loss rates to within 1\%. We include all incoming partial waves $L_\textrm{in}\le L_\textrm{max}$ in the summations used to evaluate cross sections.

\begin{figure}[tbp]
\centering
\includegraphics[width=0.45\textwidth]{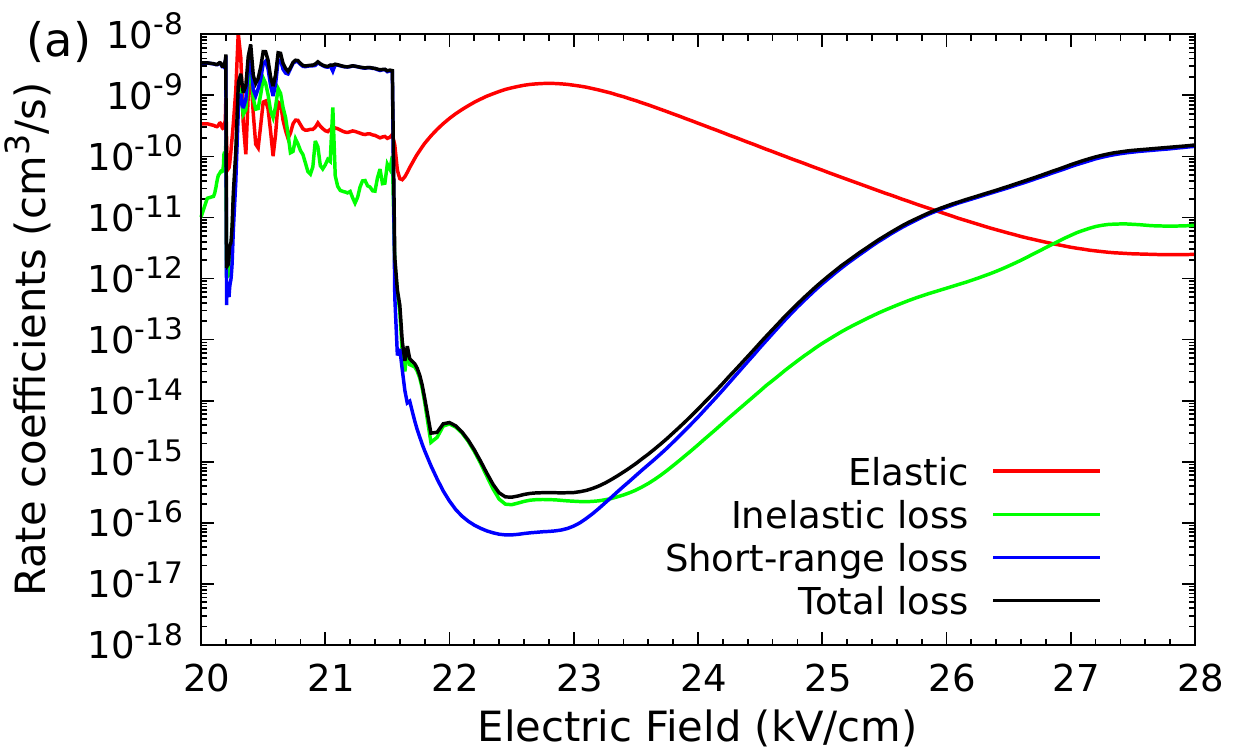}
\includegraphics[width=0.45\textwidth]{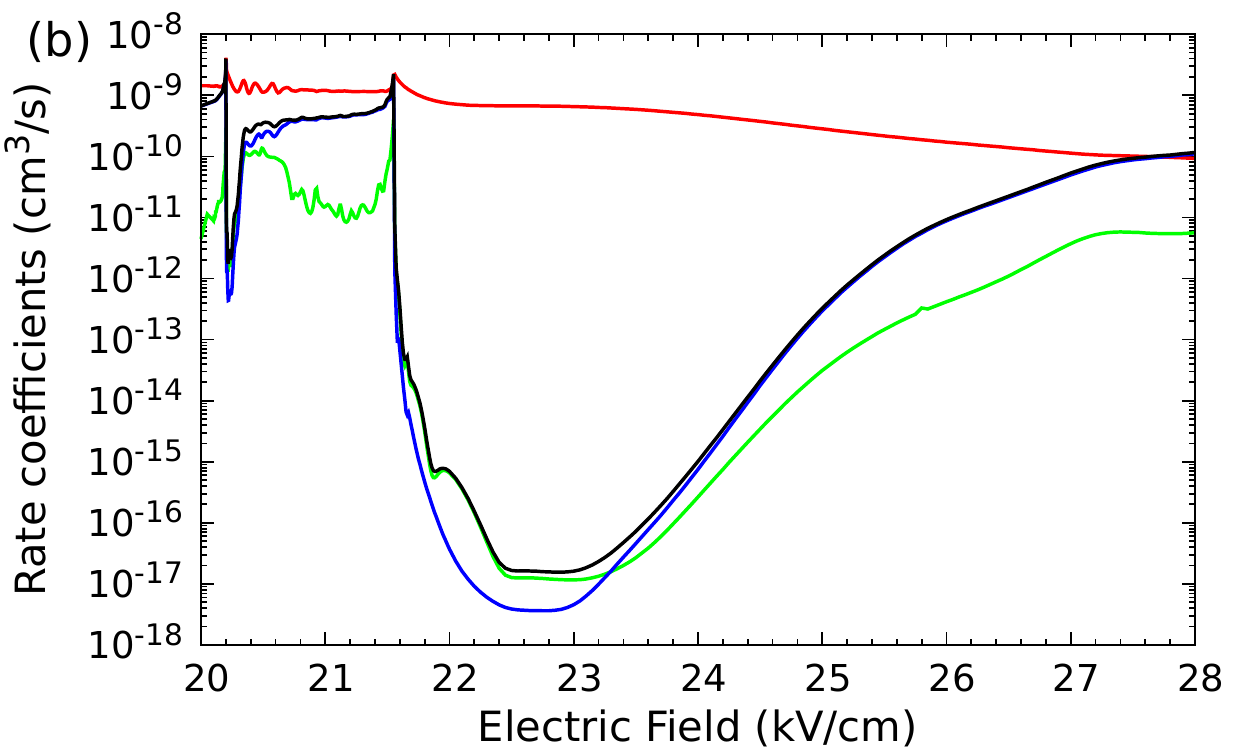}
\caption{Rate coefficients for spin-free CaF elastic collisions and loss processes as a function of electric field for (a) $E_\textrm{coll}/k_\textrm{B}=10$\ nK and (b) $E_\textrm{coll}/k_\textrm{B}=10\ \mu$K. The range of electric fields spans the crossings of (1,0)+(1,0) with (0,0)+(2,0) at 21.55 kV/cm and with (0,0)+(2,$\pm1$) at 20.20 kV/cm. The calculations use the large rotor basis set with $L_\textrm{max}=20$.
}
\label{fig:rate_coeff}
\end{figure}

\begin{figure}[tbp]
\centering
\includegraphics[width=0.45\textwidth]{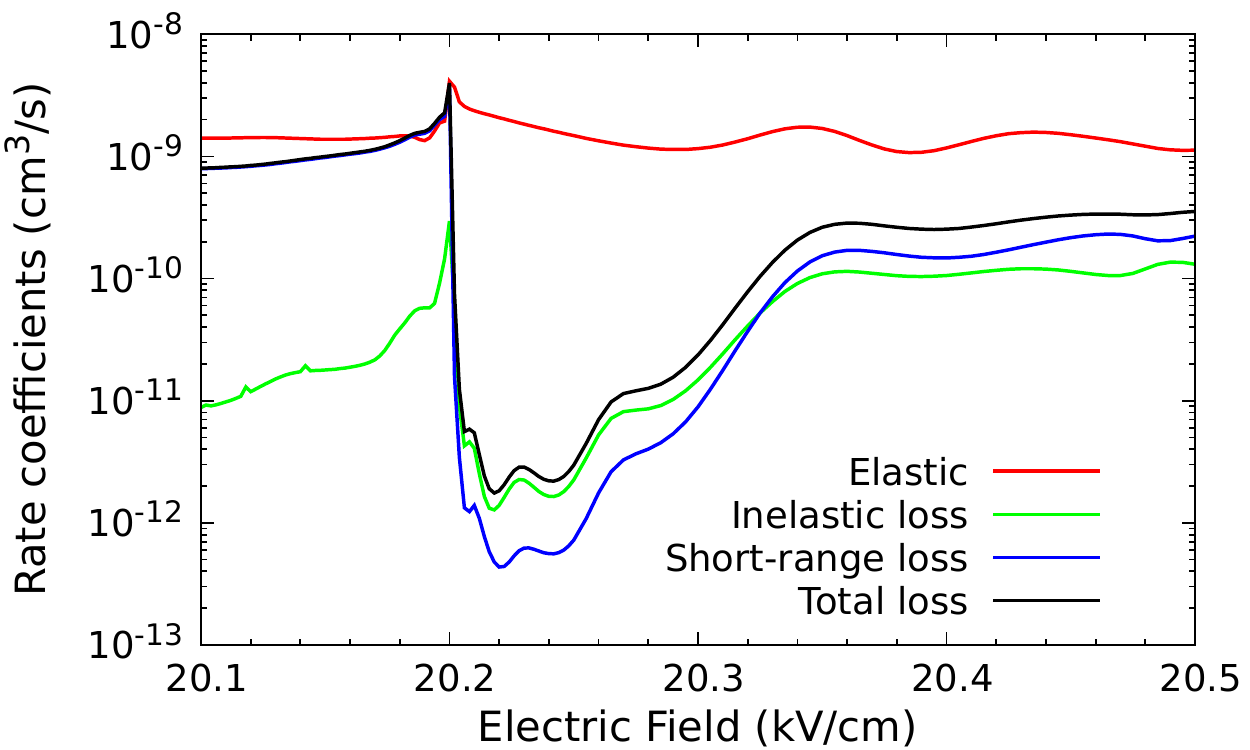}
\caption{Rate coefficients for spin-free CaF elastic collisions and loss processes as a function of electric field for $E_\textrm{coll}/k_\textrm{B}=10\ \mu$K at fields near the crossing between (1,0)+(1,0) and (0,0)+(2,$\pm1$). The calculations use the large rotor basis set with $L_\textrm{max}=20$.
}
\label{fig:rate_coeff_0021}
\end{figure}

Figure~\ref{fig:rate_coeff} shows the spin-free rate coefficients for elastic collisions and loss processes of CaF molecules initially in ($\tilde{n},m_n)$ = (1,0), as a function of electric field, in the vicinity of the crossings at 21.55 and 20.20 kV/cm. We note that the inelastic loss shown here includes only processes that occur outside $R_\textrm{absorb}=50\ a_0$, and any inelastic loss that occurs inside this distance is counted as short-range loss. The cross sections for both inelastic and short-range loss are suppressed dramatically over a wide range of fields above the crossing with (0,0)+(2,0) at 21.55 kV/cm, with a minimum near 23 kV/cm. The elastic cross section at 10 nK shows a large oscillation and enhancement in this field range, due principally to the variation in s-wave scattering length described below. At 10 $\mu$K, however, the elastic scattering is dominated by higher partial waves and this feature is absent.

The calculated ratio of elastic to inelastic rate coefficients at 23 kV/cm is about seven orders of magnitude at $E_\textrm{coll}/k_\textrm{B}=10\,\textrm{nK}$ and eight orders of magnitude at $10\,\mu\textrm{K}$. The very large value of the elastic rate coefficient will make evaporative cooling efficient. For example, a sample in a typical crossed optical dipole trap with an initial density of $10^{11}$ cm$^{-3}$ and temperature of 5 $\mu$K can be evaporated to BEC in a few seconds. With a BEC density of order $10^{13}$ cm$^{-3}$ and a rate coefficient for loss below $10^{-15}$ cm$^{3}$~s$^{-1}$, the collisional limit to the lifetime exceeds 100~s. Since collisional loss is suppressed over quite a wide range of fields, significant tuning of the dipole moment is achievable.

Figure \ref{fig:rate_coeff_0021} shows an expanded view of the rate coefficients near the lower-field crossing, between (1,0)+(1,0) and (0,0)+(2,$\pm1$). The loss rates are suppressed in this region too, but not as strongly and over a much narrower range of fields. There is no significant feature near 18.3 kV/cm, where (1,0)+(1,0) crosses (0,0)+(2,$\pm2$); these pair states are not directly coupled by $\hat{H}_\textrm{dd}$, which can change $m_n$ only by 0 or $\pm1$.

\begin{figure}[tbp]
\begin{center}
\includegraphics[width=\figwidth,clip]{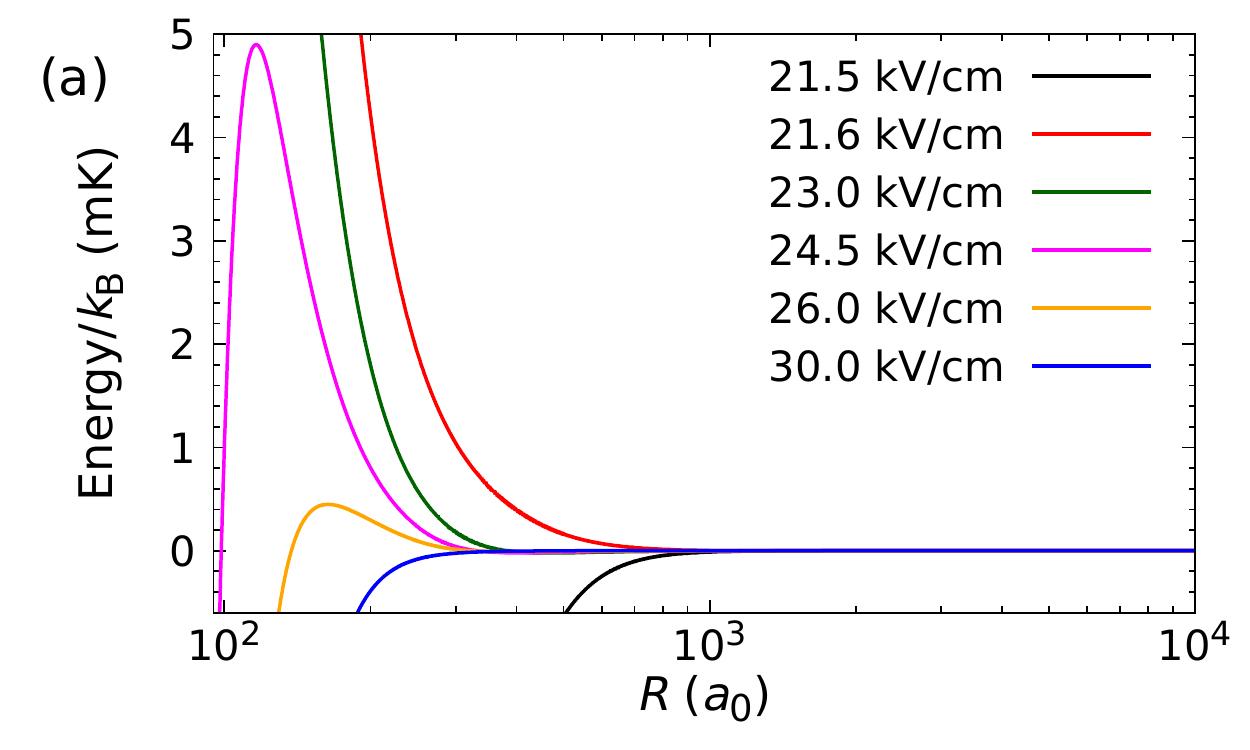}
\includegraphics[width=\figwidth,clip]{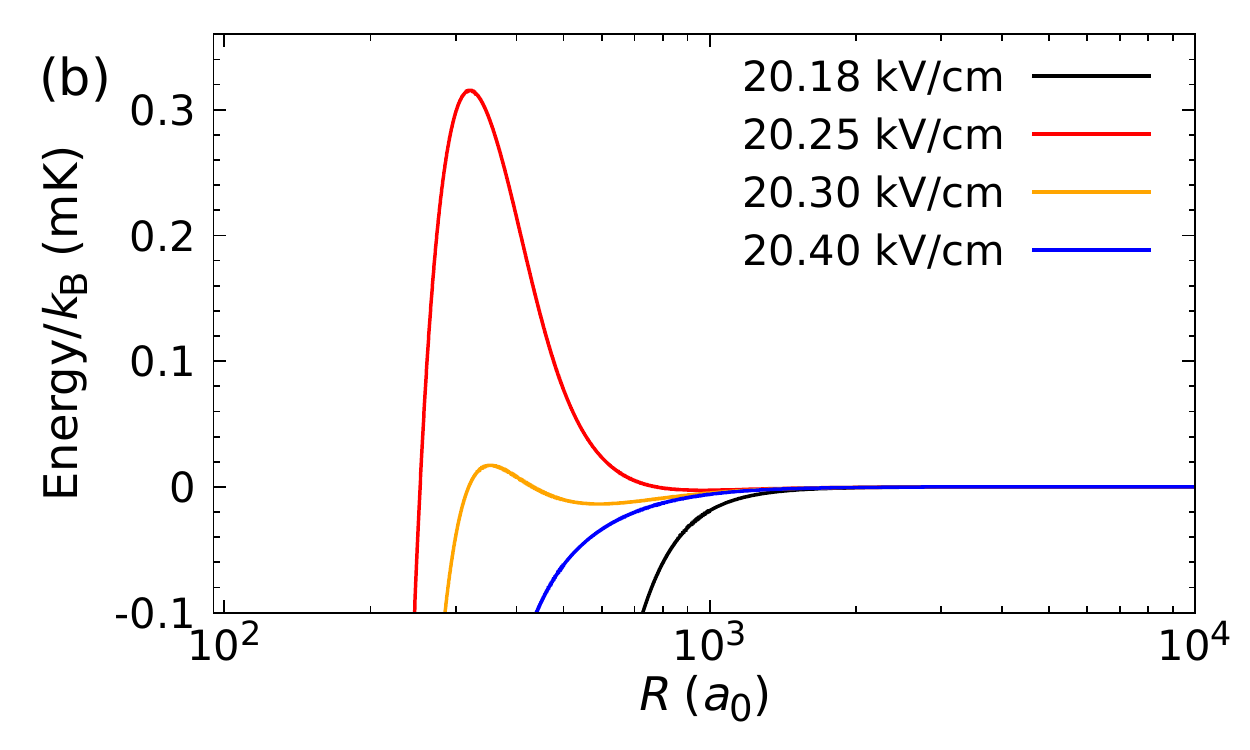}
\caption{Adiabat for the incoming channel that corresponds to (1,0)+(1,0) and $L=0$ at long range, for a variety of electric fields relevant to shielding near (a) the crossing between (1,0)+(1,0) and (0,0)+(2,0) at 21.55 kV/cm and (b) the crossing between (1,0)+(1,0) and (0,0)+(2,$\pm$1) at 20.20 kV/cm.
}
\label{fig:adiabats1}
\end{center}
\end{figure}

The general features of the rate coefficients may be explained in terms of the adiabats. Figure~\ref{fig:adiabats1} shows the adiabat for the incoming channel that corresponds to (1,0)+(1,0) and $L=0$ at long range, for a variety of electric fields close to the crossings at 20.20 and 21.55 kV/cm. At fields slightly above each crossing, the adiabats are repulsive at distances of a few hundred bohr due to mixing with the lower threshold; it is this repulsion that can prevent molecules reaching short range and produce shielding. At fields below the crossing, the same mixing causes attraction, so there is no shielding.

The adiabats also explain the differences in depth and width between the two shielding features. For the feature above 21.55 kV/cm, due to the crossing with (0,0)+(2,0), there is a substantial barrier in the incoming channel for $L=0$. The repulsive outer limb of this barrier is due principally to mixing with (0,0)+(2,0), but this competes with attractive rotational dispersion interactions due to pair states further away. It is the rotational dispersion that limits the height of the barrier. However, because of the large rotational constant and small dipole-dipole energy scale of CaF \cite{Gonzalez-Martinez:adim:2017}, rotational dispersion is relatively weaker here than for KRb \cite{Wang:dipolar:2015}, so the barrier is higher and exists over a wider range of field. By contrast, for the feature above 20.20 kV/cm, due to crossing with (0,0)+(2,$\pm1$), the barrier is limited by interactions with (0,0)+(2,0), which lies only $70\ \textrm{mK} \times k_\textrm{B}$ higher than (1,0)+(1,0) at 20.20 kV/cm and quickly comes closer as the field increases towards 21.55 kV/cm. This gives a much smaller barrier, and correspondingly weaker shielding, which extends over only a narrow range of fields. The overall result is that the feature above 21.55 kV/cm is much more pronounced and extends over a much wider range of fields than in KRb and similar systems, while the feature above 20.20 kV/cm remains relatively weak and narrow. The effect of the rotational constant on shielding above the higher-field crossing has previously been discussed by Gonz\'alez-Mart\'inez \emph{et al.}\ \cite{Gonzalez-Martinez:adim:2017}, although they did not interpret the effect in terms the adiabats.

\subsection{Dependence on collision energy}
\label{sec:Ecoll}

In the absence of long-range anisotropy, low-energy scattering is usually dominated by s-wave collisions, with incoming $L_\textrm{in}=0$. The diagonal S-matrix element in the incoming channel, $S_{00}(k_0)$, may be characterized by a complex energy-dependent scattering length $a(k_0)$,
\begin{equation}
a(k_0) = \alpha(k_0)-i\beta(k_0) = \frac{1}{ik_0} \left(\frac{1-S_{00}(k_0)}{1+S_{00}(k_0)}\right),
\label{eq:a-k}
\end{equation}
where $k_0=(2\mu E_\textrm{coll}/\hbar^2)^{1/2}$.
The corresponding contribution to the elastic scattering cross section is \cite{Hutson:res:2007}
\begin{equation}
\sigma_\textrm{el,00}(k_0) = \frac{4\pi g|a|^2}{1+k_0^2|a|^2+2k_0\beta},
\label{eq:sigma-el-00}
\end{equation}
where $g=2$ for identical bosons. This is not the complete s-wave contribution to the elastic cross section, because it neglects contributions from $L$-changing collisions with $L_\textrm{out}>0$.
We refer to $\sigma_\textrm{el,00}(k_0)$ as the diagonal s-wave contribution to the cross section. Similarly, the expression that is commonly used for the inelastic cross section,
\begin{equation}
\sigma_\textrm{inel,00}(k_0) = \frac{4\pi g\beta}{k_0(1+k_0^2|a|^2+2k_0\beta)},
\label{eq:sigma-loss-00}
\end{equation}
actually includes contributions from $L$-changing collisions that form part of the elastic cross section. These distinctions are often not important for atomic collisions, but they \emph{are} important in the present case.

\begin{figure}[tbp]
\centering
\includegraphics[width=0.48\textwidth]{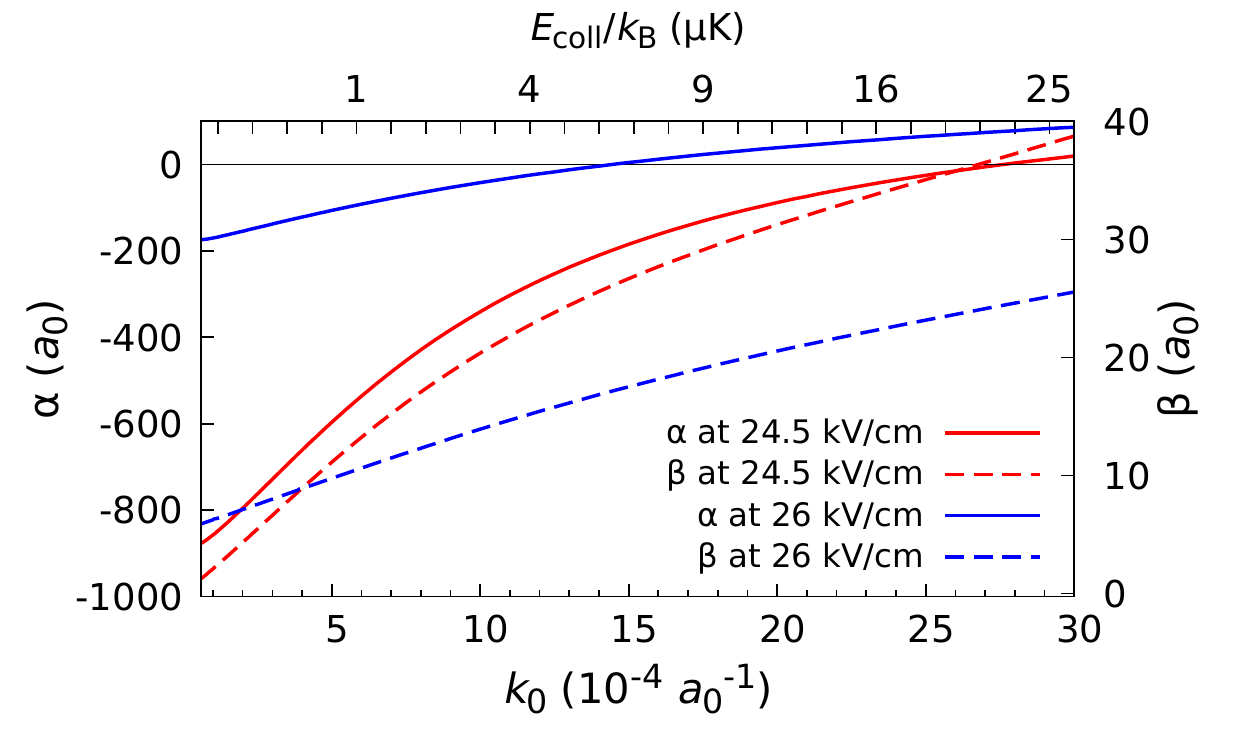}
\caption{Real (solid) and imaginary (dashed) parts of scattering length $a(k_0)$ as a function of incident wave vector $k_0$ at $24.5~$kV/cm (red) and $26~$kV/cm (blue). The corresponding collision energies are shown on the upper horizontal axis; this axis is linear in $\sqrt{E_\textrm{coll}/k_\textrm{B}}$, with tick marks separated by $0.2\ \sqrt{\mu\text{K}}$.
}
\label{fig:scatlen}
\end{figure}

Figure \ref{fig:scatlen} shows the calculated real and imaginary parts of the scattering length as a function of $k_0$ from spin-free calculations at fields of 24.5 and 26 kV/cm. At both fields $\alpha(k_0)$ is large and negative at low energy, but its magnitude decreases substantially as the energy increases. The negative low-energy scattering length arises because the lowest adiabat (correlating with $L=0$) is attractive and behaves as $-C_4R^{-4}$ at long range \cite{Karman:dipole:2018}. This occurs because there are off-diagonal matrix elements of $\hat{H}_\textrm{dd}$ between $L=0$ and 2. As a result, a long-range well exists outside the shielding barrier. This well is not deep enough to support a bound state for CaF in the range of fields considered here, so produces a scattering length that is negative at low energy.

For atom-atom scattering with a long-range potential of the form $-C_6R^{-6}$, both $\alpha(k_0)$ and $\beta(k_0)$ are independent of $k_0$ at limitingly low energy, with a leading correction term proportional to $k_0^2$ \cite{Hinckelmann:1971}. For a potential of the form $-C_4R^{-4}$, however, the leading correction to $\alpha(k_0)$ is linear in $k_0$. This behavior is seen in Fig.\ \ref{fig:scatlen}. $\alpha(k_0)$ crosses zero near $E_\textrm{coll}/k_\textrm{B}=22\ \mu$K at 24.5 kV/cm and near $6\ \mu$K at 26 kV/cm, producing corresponding minima in $\sigma_\textrm{el,00}(k_0)$ as a function of energy. The imaginary part $\beta(k_0)$ also varies linearly with $k_0$ at low energies, due to the elastic contribution from $L$-changing collisions, and will be described below.

\begin{figure}[tbp]
\centering
\includegraphics[width=0.45\textwidth]{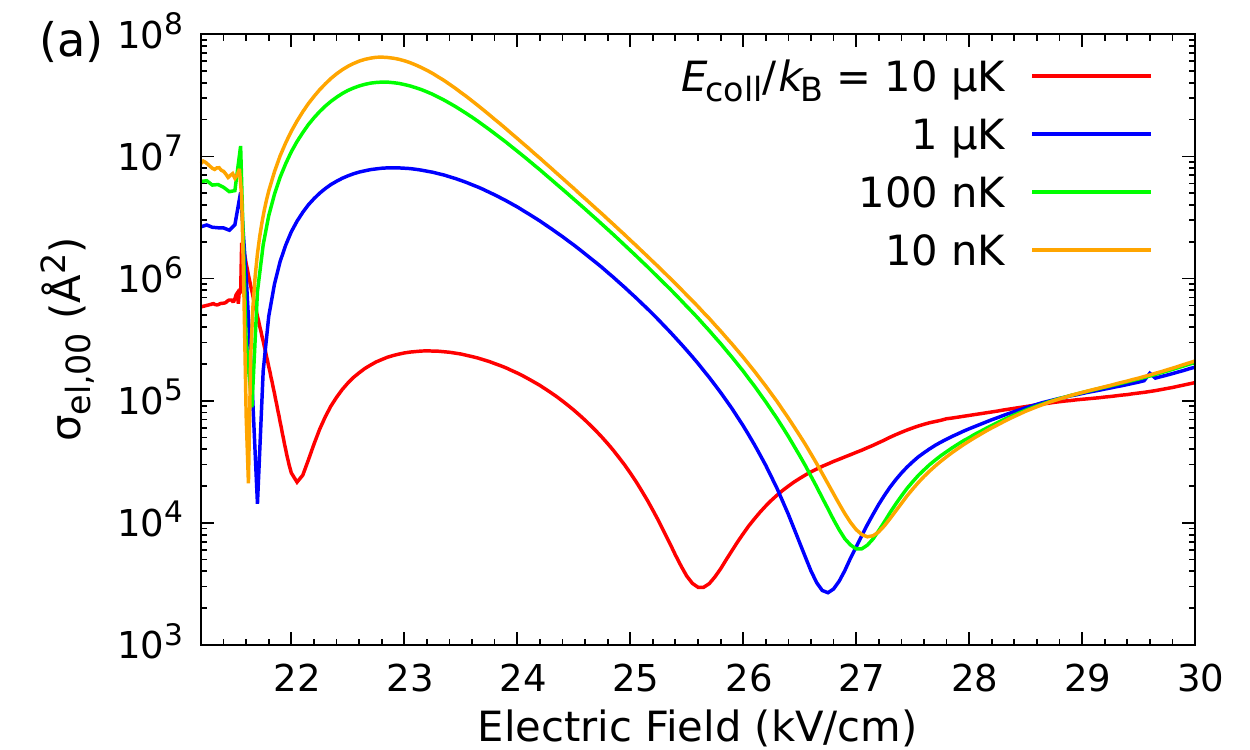}
\includegraphics[width=0.45\textwidth]{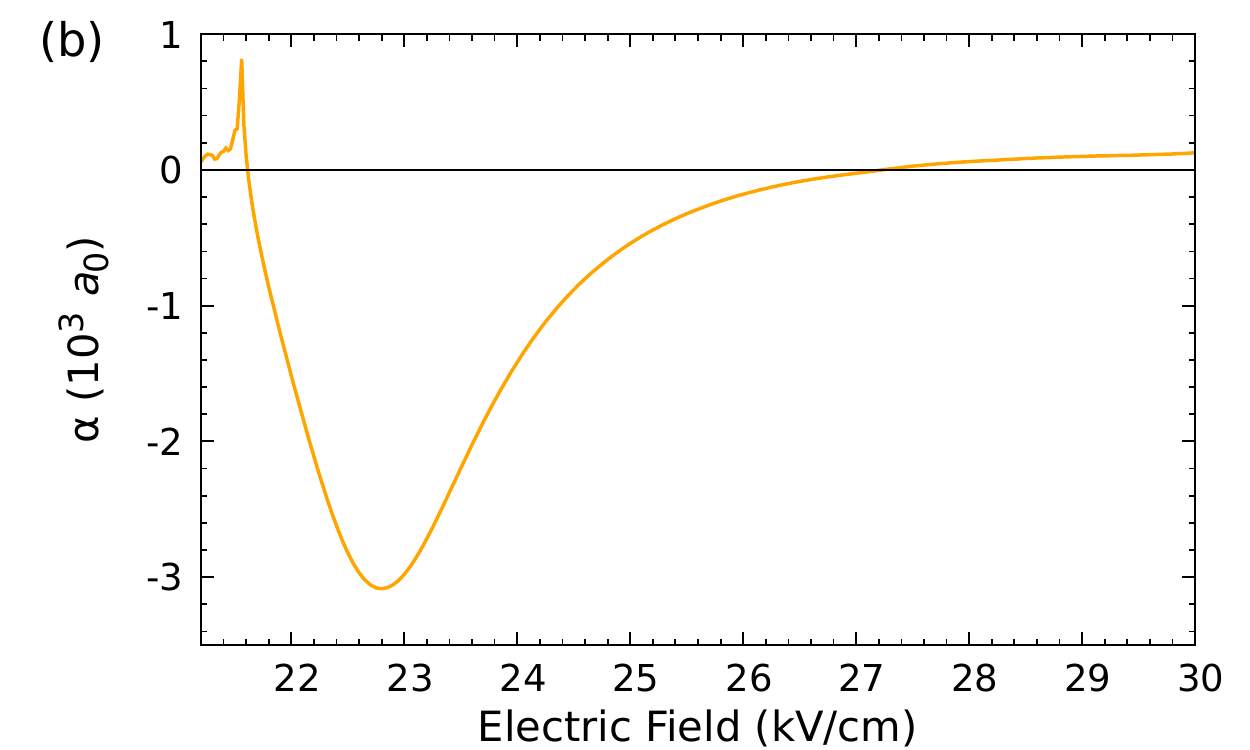}
\includegraphics[width=0.45\textwidth]{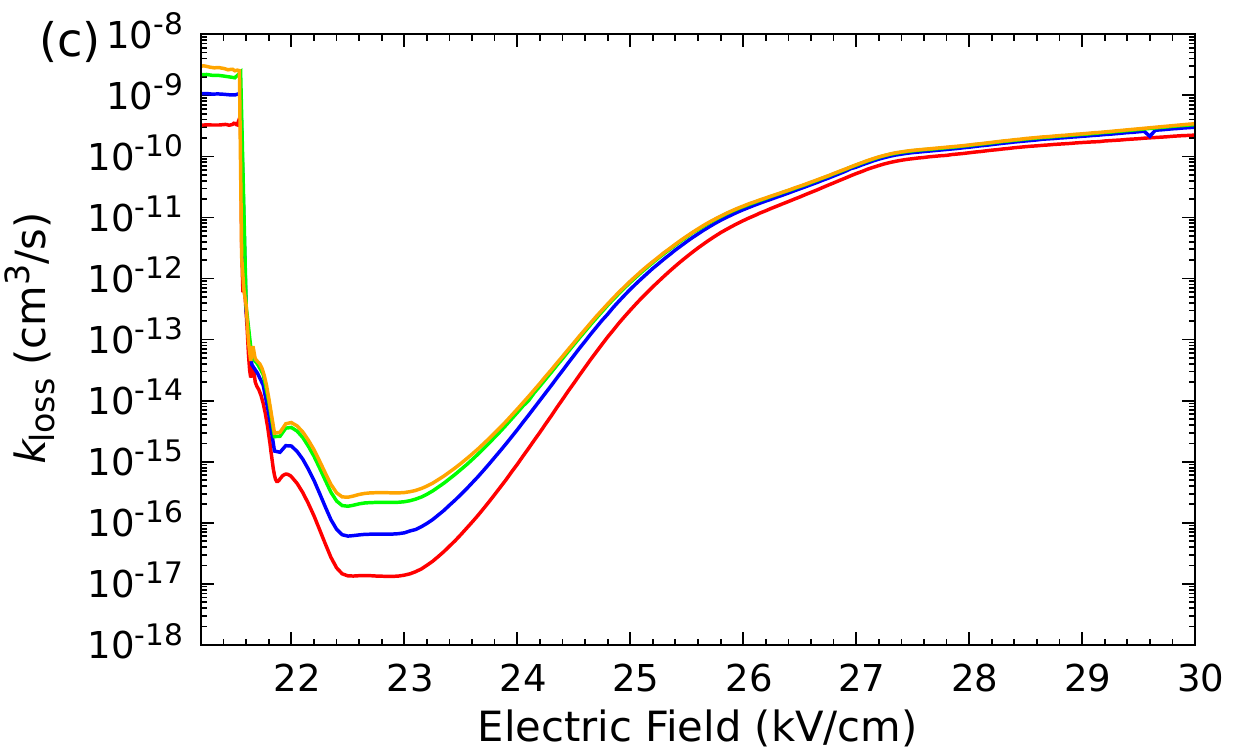}
\caption{(a) Diagonal s-wave contribution to the elastic cross sections;
(b) Real part of low-energy scattering length $\alpha$, calculated at $E_\textrm{coll}/k_\textrm{B}=10$~nK.
(c) s-wave contribution to rate coefficient for total loss, $k_\textrm{loss}$, with $L_\textrm{in}=0$ but summed over $L_\textrm{out}$.
}
\label{fig:s-energy-dependence}
\end{figure}

Figure \ref{fig:s-energy-dependence}(a) shows $\sigma_\textrm{el,00}(k_0)$ as a function of electric field for several collision energies. The cross section at 10~nK directly reflects the field-dependence of $a(0)$, whose real part $\alpha(0)$ crosses zero near 21.6 kV/cm and again near 27.2 kV/cm, as shown in Fig.\ \ref{fig:s-energy-dependence}(b). The higher-field crossing and the corresponding minimum in $\sigma_\textrm{el,00}(k_0)$ move to lower field at higher energies as $\alpha(0)$ becomes more negative. The lower-field crossing moves in the opposite direction. Figure \ref{fig:s-energy-dependence}(c) shows the s-wave contributions to rate coefficients for total loss for various collision energies. In this case the energy dependence is much simpler, with a slow but steady drop in cross section as energy increases.

\begin{figure}[tbp]
\centering
\includegraphics[width=0.45\textwidth]{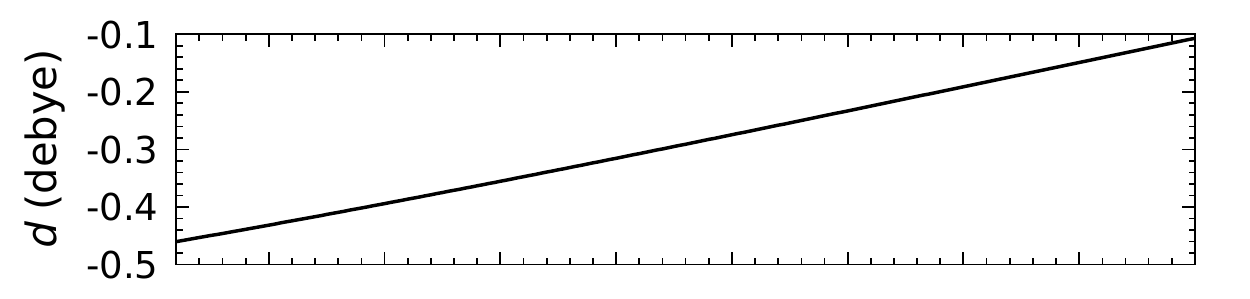}
\includegraphics[width=0.45\textwidth]{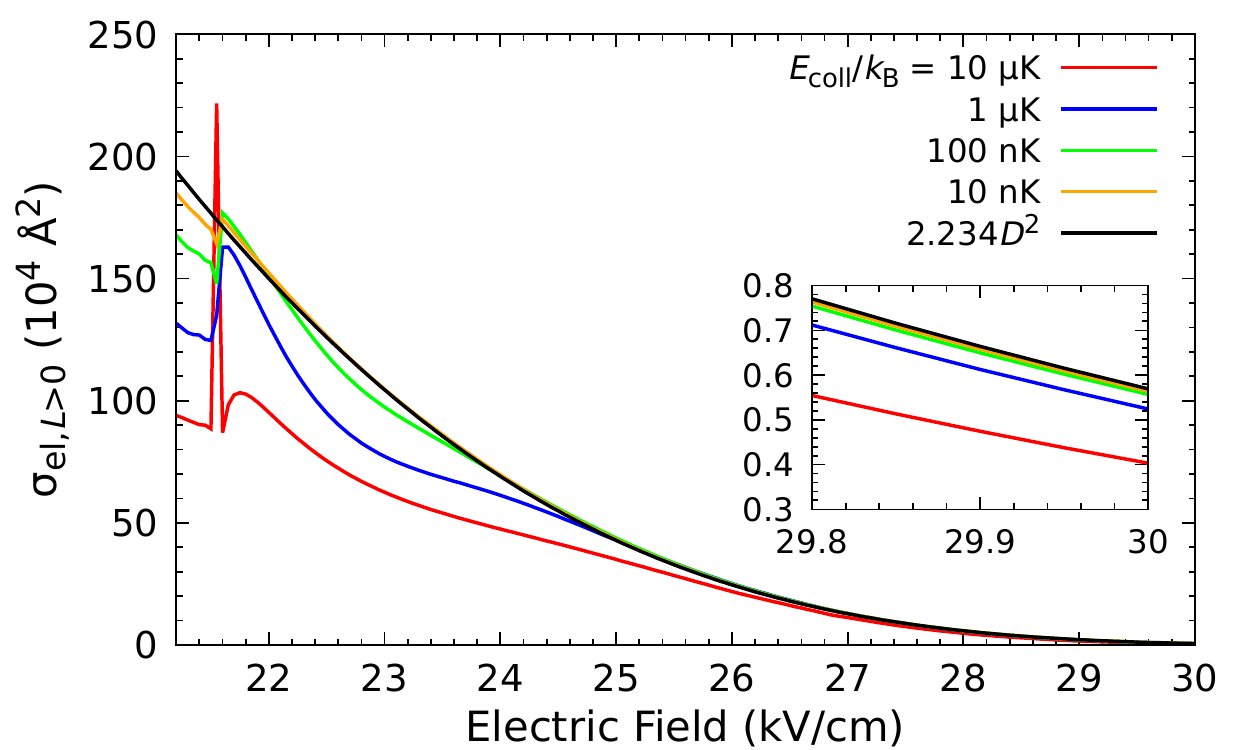}
\caption{Higher-$L$ contributions $\sigma_{\textrm{el},L{>}0}$ to elastic cross sections for various collision energies, compared with the Born approximation $2.234D^2$. These calculations use a minimal rotor basis set with only the pair levels (1,0)+(1,0) and (0,0)+(2,0) in class 1, with $L_\textrm{max}=20$; this is well converged for elastic cross sections in this range of fields. The upper panel shows the space-fixed dipole moment $d$ of the state ($\tilde{n},m_n)=(1,0)$.
}
\label{fig:bct}
\end{figure}

\begin{figure}[tbp]
\centering
\includegraphics[width=0.45\textwidth]{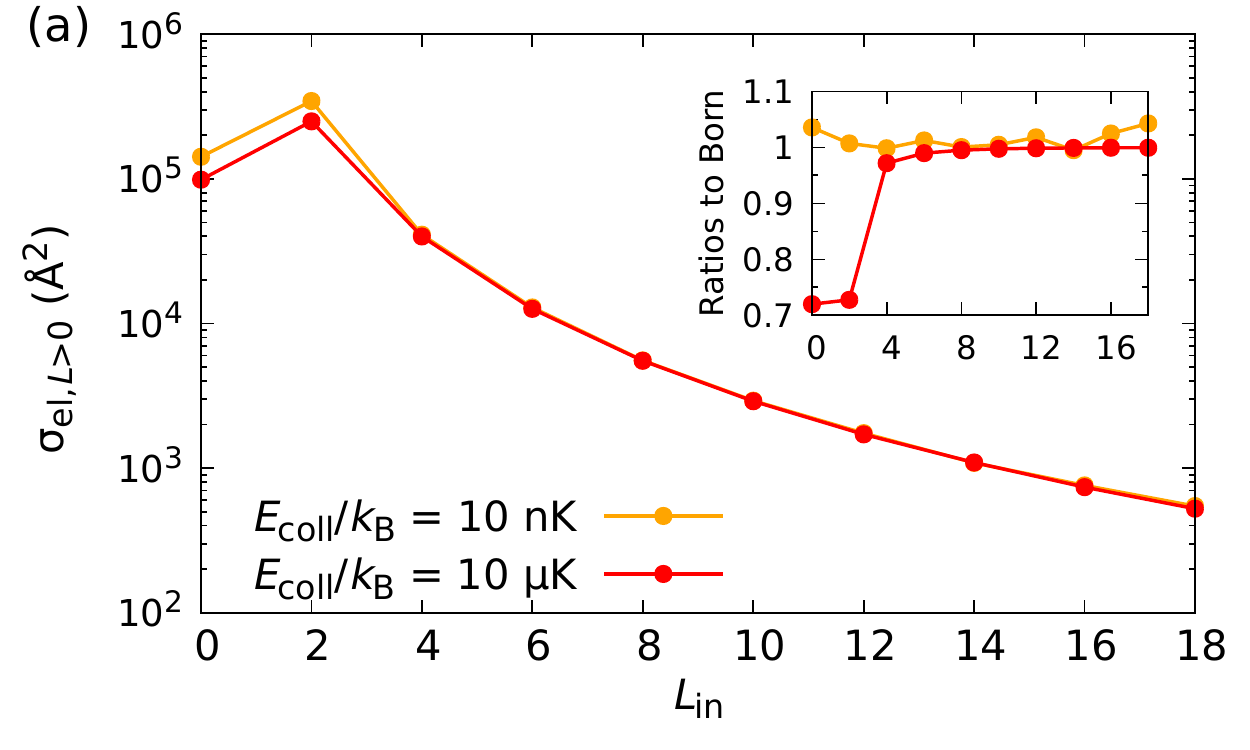}
\includegraphics[width=0.45\textwidth]{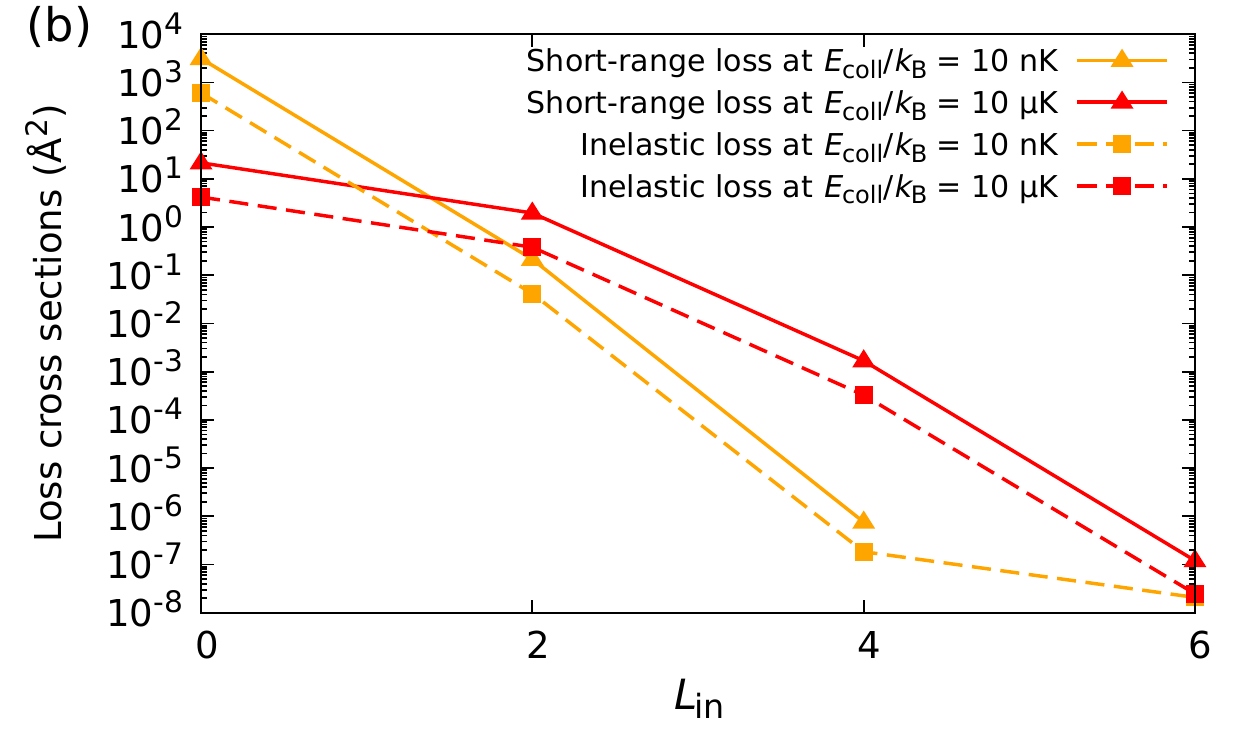}
\caption{Contributions of individual incoming partial waves to cross sections at a field of 24.5 kV/cm for (a) elastic cross section, excluding the diagonal s-wave contribution; (b) inelastic and short-range loss cross sections. The inset in panel (a) shows the ratios of the coupled-channel cross sections to values from the Born approximation.
}
\label{fig:Lin-contribution}
\end{figure}

\begin{figure}[tbp]
\centering
\includegraphics[width=0.45\textwidth]{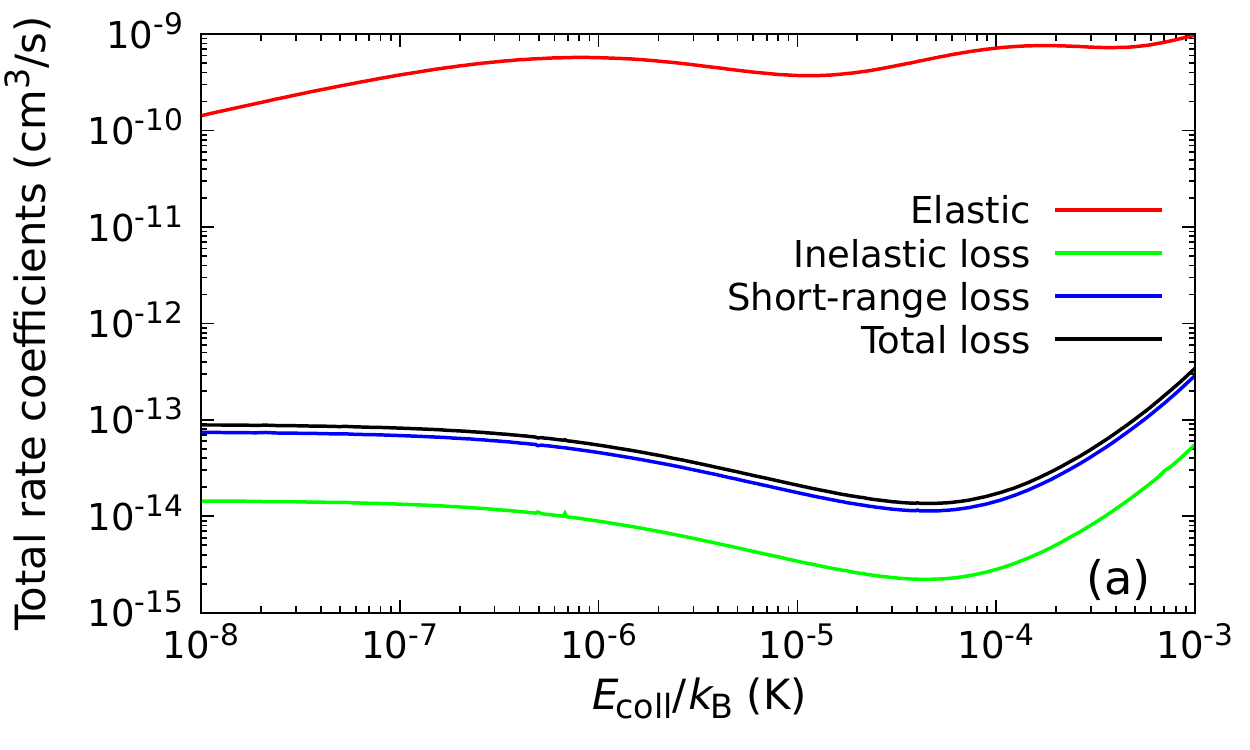}
\includegraphics[width=0.45\textwidth]{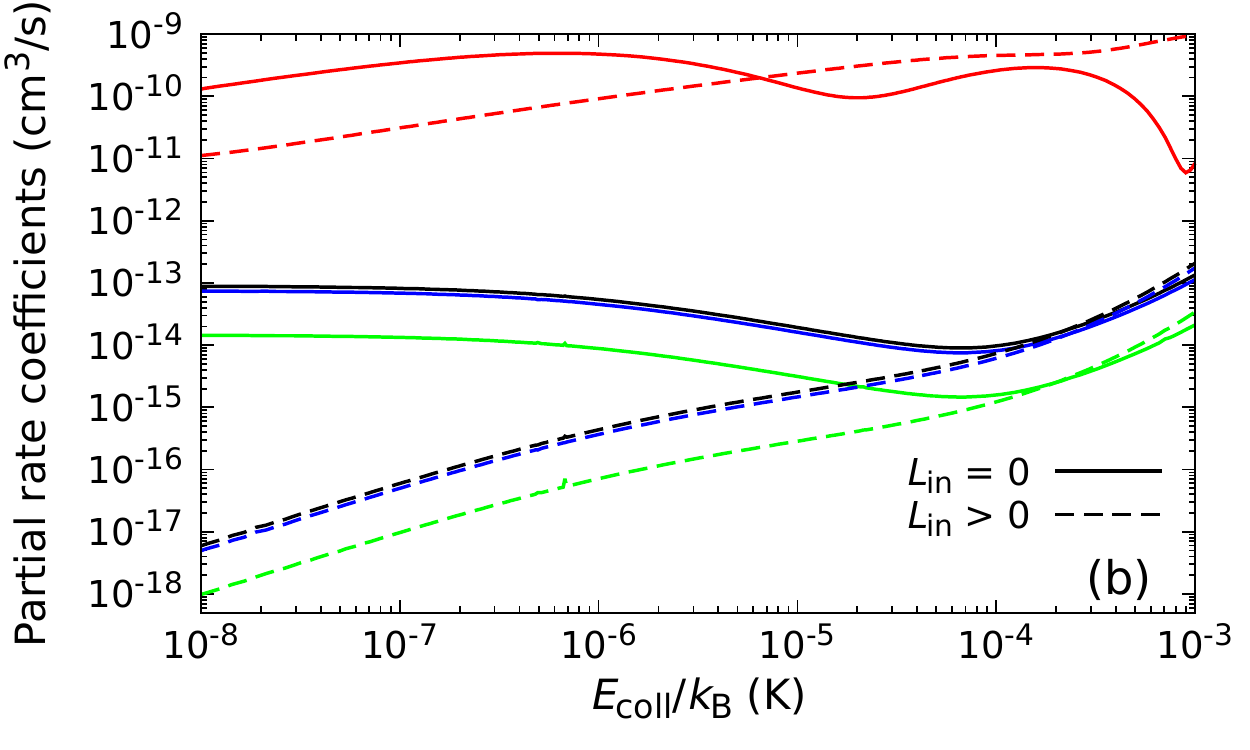}
\caption{Rate coefficients as a function of collision energy at a field of $24.5~$kV/cm. Panel (a) shows the rate coefficients summed over $L_\textrm{in}$, whereas panel (b) shows them separated into contributions from $L_{\rm in}=0$ (solid lines) and $L_{\rm in}>0$ (dashed lines).
}
\label{fig:rates-enerdepen}
\end{figure}

For dipole-dipole scattering, higher partial waves also play an important role. The dipole-dipole interaction couples different partial waves $L$, and dies off only slightly faster ($R^{-3}$) than the centrifugal separation between the channels ($R^{-2}$). Because of this, there are substantial contributions to elastic cross sections from $L_\textrm{in}>0$ and/or $L_\textrm{out}>0$, which we refer to as $\sigma_{\textrm{el},L{>}0}$. For dipoles fixed in space, the contributions to elastic cross sections may be estimated from a Born approximation \cite{Bohn:BCT:2009}, and for identical bosons at limitingly low energy they sum to $2.234D^2$, where $D=d_1 d_2 \mu/(4\pi\epsilon_0\hbar^2)$ is the dipole length and $d_k$ are the space-fixed dipoles induced by the electric field. Of this, $1.396D^2$ arises from $M_\textrm{tot}=0$.

Figure \ref{fig:bct} shows the higher-$L$ contributions to elastic cross sections as a function of field for collision energies between $10~\text{nK} \times k_{\rm B}$ and $10~\mu\text{K} \times k_{\rm B}$. It also shows the Born approximation $2.234D^2$, which varies with field because of the variation in the induced dipole shown in Fig.\ \ref{fig:bct}. The Born approximation is very accurate at 10~nK, but breaks down substantially at higher energies, particularly at fields in the range important for shielding. Figure \ref{fig:Lin-contribution}(a) shows the higher-$L$ contribution for a single field of 24.5 kV/cm, further broken down into contributions from individual values of $L_\textrm{in}$. It may be seen that the breakdown of the Born approximation occurs principally for $L_\textrm{in}<4$. Thus, contributions from $L_\textrm{in}>L_\textrm{max}$ and $L_\textrm{out}>L_\textrm{max}$, which are not captured by our coupled-channel calculations, can instead be accounted for with good accuracy using the Born approximation.

The higher-$L$ contributions to elastic cross sections are determined by dipole-dipole interactions at very long range, typically $R>500\ a_0$. The diagonal s-wave contribution arises from shorter-range physics, governed by the outer turning point of the shielding barrier in the adiabat for $L=0$, but still at $R>200\ a_0$ for fields where shielding is effective. At these distances the Hamiltonian is strongly dominated by the dipole-dipole terms, so we expect the elastic rate coefficients obtained here to be quantitatively predictive.

Figure \ref{fig:Lin-contribution}(b) shows the breakdown of the cross sections for inelastic and short-range loss into contributions from individual values of $L_\textrm{in}$. The losses from $L_\textrm{in}>4$ are very small, and partial waves $L_\textrm{in}>6$ contribute less than one part in $10^8$ to the cross sections.

In the Born approximation, the off-diagonal S-matrix elements for $L$-changing elastic collisions are proportional to $k_0$ at low energy \cite{Bohn:BCT:2009} \footnote{Equation \ (14) of Ref.\ \cite{Bohn:BCT:2009} incorrectly includes factors of 32 that should be 4.}.
At limitingly low energy,
\begin{equation}
\beta(k_0) = \beta_\textrm{loss} + (1/45) D^2 k_0,
\label{eq:beta-linear}
\end{equation}
where only $\beta_\textrm{loss}$ represents loss and the term linear in $k_0$ results from elastic scattering. This is the behavior seen in Fig.\ \ref{fig:scatlen}. It may be noted that Ref.\ \cite{Gonzalez-Martinez:adim:2017} calculated loss rates using Eq.\ (\ref{eq:sigma-loss-00}), with $\beta(k_0)$ obtained from coupled-channel calculations of $S_{00}$ at $E_\textrm{coll}/k_\textrm{B} = 100$~nK. This procedure can dramatically overestimate loss rates where shielding is most effective; for CaF at 23 kV/cm, $\beta_\textrm{loss} = 1.1\times 10^{-4}\ a_0$
but $\beta(k_0) = 7.7\ a_0$ at $E_\textrm{coll}/k_\textrm{B}=100$~nK, so it would overestimate the loss rate by about a factor of $10^5$. The absolute loss rate is important because it will determine the lifetime of the ultracold dipolar gas.

Figure \ref{fig:rates-enerdepen}(a) shows the dependence of the total rate coefficients on energy for a field of 24.5 kV/cm. Figure \ref{fig:rates-enerdepen}(b) breaks these down into contributions from $L_\textrm{in}=0$ and $L_\textrm{in}>0$.  For the elastic rates, the two contributions have the same threshold law. Nevertheless, $L_\textrm{in}=0$ dominates at energies below 10~$\mu\textrm{K} \times k_\textrm{B}$, while $L_\textrm{in}>0$ dominates above that. For the inelastic and short-range loss, the contributions from $L_\textrm{in}=0$ and $L_\textrm{in}>0$ have different threshold laws: the s-wave contribution to the rate coefficient is independent of energy at very low energy, while that from $L_\textrm{in}>0$ is proportional to $E_\textrm{coll}$. The two contributions are comparable above 10~$\mu$K. This explains the very different dependence of $k_\textrm{el}$ on field at $E_\textrm{coll}/k_\textrm{B} = 10$~nK and 10~$\mu$K in Fig.\ \ref{fig:rate_coeff}.

\subsection{Effects of electron and nuclear spins}
\label{sec:spins}

Qu\'em\'ener \emph{et al.}\ \cite{Quemener:2016} have argued that the spin quantum numbers should behave as spectators during shielding collisions, so that spin-free calculations are adequate. Nevertheless, they presented results that indicate that, for fields even slightly (10\%) above the optimum field for shielding in RbSr, inclusion of the full spin structure can enhance shielding by a factor of 10, increasing to $10^4$ at fields 30\% higher. They attributed this effect to additional repulsion due to spin states neglected in the spin-free calculations. We have therefore carried out a detailed investigation of the effects of spin on shielding for CaF.

\begin{figure}[tbp]
   \includegraphics[width=0.5\textwidth]{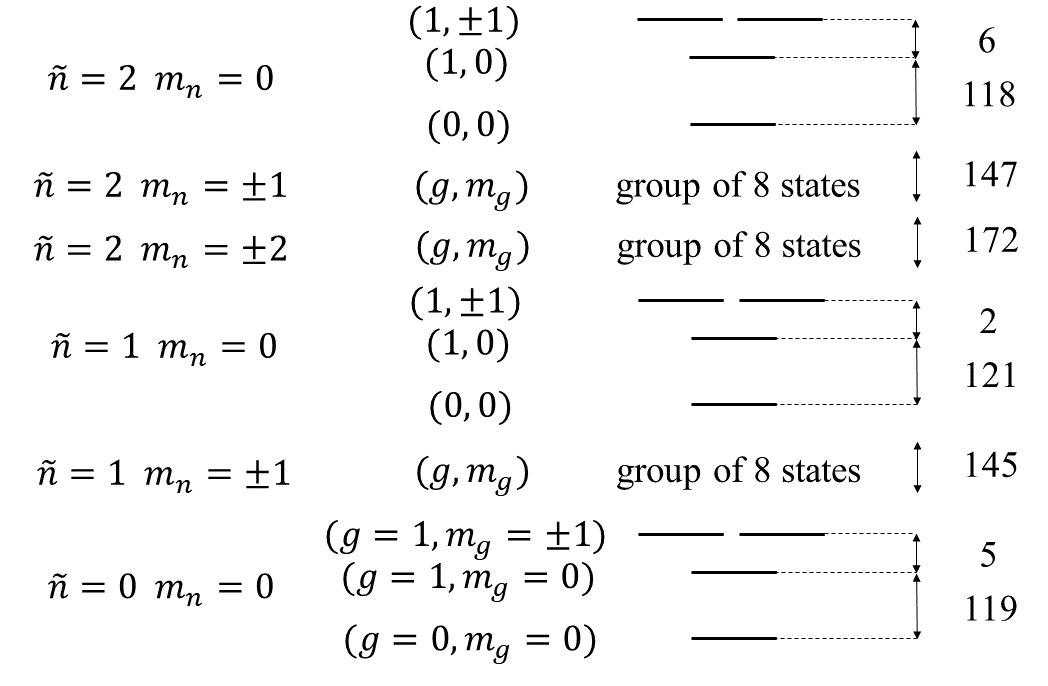}
    \caption{Hyperfine splittings of CaF rotor levels (in MHz) at electric field $23~$kV/cm.
}
\label{fig:energylevels}
\end{figure}

The effects of electron and nuclear spins on the individual molecules are described by the Hamiltonian for fine and hyperfine structure,
\begin{equation}
\hat{h}_{\mathrm{fhf}} = \gamma\boldsymbol{\hat{s}} \cdot \boldsymbol{\hat{n}} + \zeta_\mathrm{F} \boldsymbol{\hat{i}} \cdot \boldsymbol{\hat{s}} + t\sqrt{6}T^2(C) \cdot T^2(\boldsymbol{\hat{i}},\boldsymbol{\hat{s}}) + c_\mathrm{F} \boldsymbol{\hat{i}} \cdot \boldsymbol{\hat{n}}.
\label{eq:ham-fhf}
\end{equation}
Here the first term represents the electron spin-rotation interaction, while the second and third terms account for the isotropic and anisotropic interactions between electron and nuclear spins. $T^2(\boldsymbol{\hat{i}},\boldsymbol{\hat{s}})$ denotes the rank-2 spherical tensor formed from $\boldsymbol{\hat{i}}$ and $\boldsymbol{\hat{s}}$, and $T^2(C)$ is a spherical tensor whose components are the Racah-normalized spherical harmonics $C^2_q(\theta,\phi)$. The last term represents the nuclear spin-rotation interaction, which is typically three orders of magnitude smaller than the others. The values of the constants $\gamma$, $\zeta_\mathrm{F}$, $t$ and $c_\mathrm{F}$ for CaF are taken from Refs.\ \cite{Childs:CaF:1981} \footnote{Reference\ \cite{Childs:CaF:1981} uses the notation of Frosch and Foley \cite{Frosch:1952}, where our $b_0$, $\gamma$, $\zeta_\textrm{F}$, $t$ and $c_\textrm{F}$ are $B$, $\gamma$, $b+c/3$, $c/3$ and $C_\textrm{I}$, respectively.}.

In the present paper, we are interested in collisions in the presence of an electric field of around 20 kV/cm. Figure~\ref{fig:energylevels} shows the fine and hyperfine splittings for the monomer states with $(\tilde{n},m_n)$ = (0,0), (1,0) and (2,0), which are the most relevant for shielding. The only fully conserved quantum number is $m_f=m_n+m_s+m_i$. For all states, however, the general pattern is that $g$, the resultant of $i$ and $s$, is approximately conserved, along with $m_n$ and $m_g$, but $m_s$ and $m_i$ are individually poorly defined.

\begin{figure}[tbp]
		\includegraphics[width=0.45\textwidth]{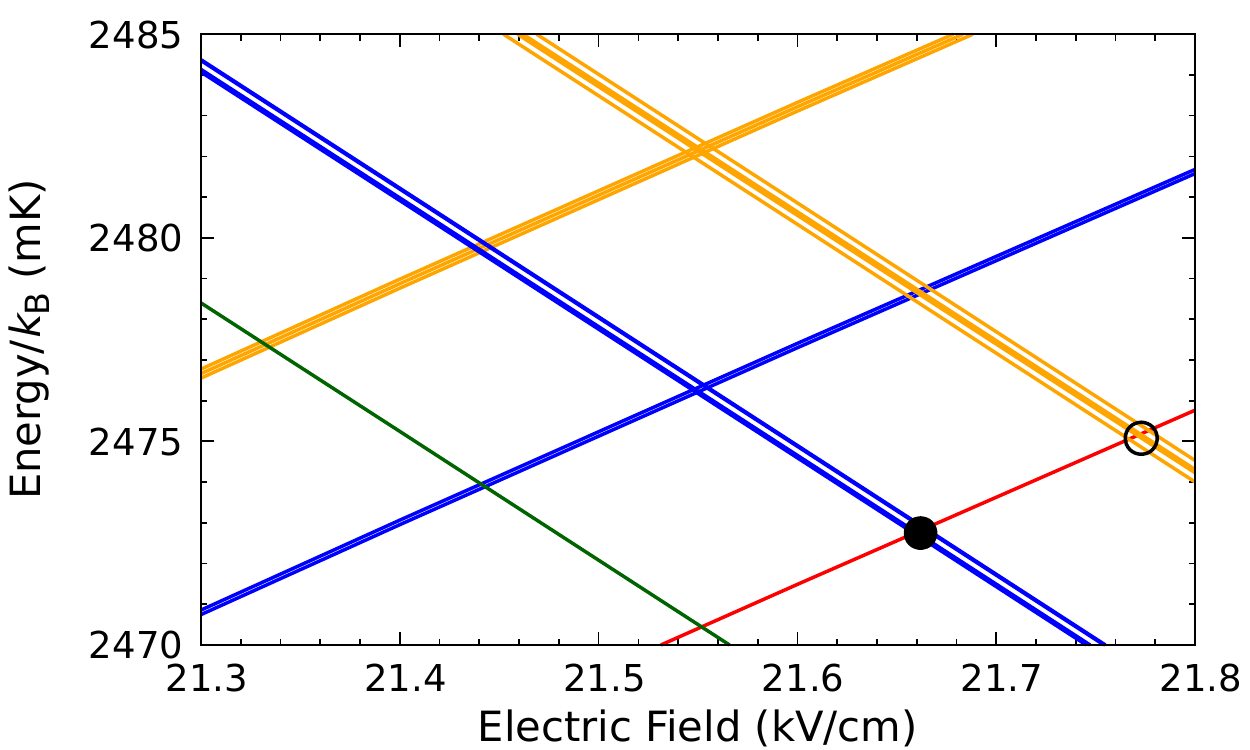}
\caption{Crossings between thresholds for CaF+CaF including spin. Crossing between states arising from (1,0)+(1,0) (sloping up) with those arising from (0,0)+(2,0) (sloping down). $g,m_g$ take all allowed values. The states with $g=0$ are color-coded as in Fig.\ \ref{fig:CaF_Stark}, while pair levels with one or both molecules excited to states with $g=1$ are shown in blue and orange, respectively. The closed and open circles indicate the crossings responsible for the spin-induced peaks in Fig.\ \ref{fig:rate_coeff-spins}.
}
\label{fig:crossings}
\end{figure}

The fine and hyperfine structure complicates the patterns of pair levels in the vicinity of crossings between spin-free levels. Figure~\ref{fig:crossings} shows the crossings of $(\tilde{n},m_n)$ = (1,0)+(1,0) with (0,0)+(2,0) when spin splittings are included. If both molecules are initially in the lowest spin component, with $(\tilde{n},m_n,g,m_g)$ = (1,0,0,0) (black line), the crossing with (0,0,0,0)+(2,0,0,0) (green line) is almost unshifted at 21.55 kV/cm, but there are additional crossings with excited spin channels at 21.66 and 21.77 kV/cm. The consequences of these are discussed below.

To solve the coupled equations for scattering, we use basis sets constructed from products of field-dressed rotor functions $|\tilde{n},m_n\rangle$, calculated without spins, and spin functions $|g,m_g\rangle$, formed as Clebsch-Gordan sums of $|s,m_s\rangle$ and $|i,m_i\rangle$.
There are four spin functions for each monomer rotor state, so 16 spin combinations for each pair state. The full pair basis set is restricted by the conservation of $M_\textrm{tot}$, which is now $m_{n1}+m_{g1}+m_{n2}+m_{g2}+M_L$, and by exchange symmetry, but the overall size of the basis set nevertheless increases by a factor of about 10 when spins are included. This increases the computer time by a factor of about 1000, so it is necessary to use smaller rotor basis sets than for spin-free calculations.

The Van Vleck transformation once again makes it possible to include the effects of well-separated basis functions without including them in the explicit basis set used to solve the coupled equations. We choose a limited set of combinations of rotor and spin functions to include in class 1. In principle, $\hat{H}_\textrm{dd}$ and every operator in Eq.\ (\ref{eq:ham-fhf}) have matrix elements connecting basis functions in class 1 with those in class 2. However, the term $\zeta_\textrm{F}\boldsymbol{\hat{i}} \cdot \boldsymbol{\hat{s}}$ is diagonal in rotor quantum numbers, and the small nuclear spin-rotation term is neglected. Terms involving spin operators in second order are independent of $R$ and have negligible effects on scattering. The terms involving $\hat{H}_\textrm{dd}$ in second order are identical to those included in the spin-free case, and are diagonal in spin quantum numbers. The additional terms that appear in a second-order Van Vleck transformation to handle spin are those first order in $\hat{H}_\textrm{dd}$ and also first-order in either the spin-rotation interaction $\hat{H}_{sn}=\gamma \sum_{k=1,2} \boldsymbol{\hat{s}}_k \cdot \boldsymbol{\hat{n}}_k$ or the anisotropic hyperfine interaction $\hat{H}_{is}^\textrm{(2)}=t \sqrt{6} \sum_{k=1,2} T^2(C_k) \cdot T^2(\boldsymbol{\hat{i}}_k,\boldsymbol{\hat{s}}_k)$. These terms are of the form
\begin{align}
\langle a | \hat{H}_{\textrm{dd},sn,\textrm{VV}} &| b \rangle \nonumber\\
= \sum_\alpha \frac{1}{2} & \Big[
\frac{\langle a | \hat{H}_\textrm{dd} | \alpha \rangle
\langle \alpha | \hat{H}_{sn} | b \rangle} {(E_a-E_\alpha)}
+ \frac{\langle a | \hat{H}_\textrm{dd} | \alpha \rangle
\langle \alpha | \hat{H}_{sn} | b \rangle} {(E_b-E_\alpha)}
\nonumber\\
+ &
\frac{\langle a | \hat{H}_{sn} | \alpha \rangle
\langle \alpha | \hat{H}_\textrm{dd} | b \rangle} {(E_a-E_\alpha)}
+ \frac{\langle a | \hat{H}_{sn} | \alpha \rangle
\langle \alpha | \hat{H}_\textrm{dd} | b \rangle} {(E_b-E_\alpha)}
\Big]
\label{eq:VV-spin}
\end{align}
and similarly for $\hat{H}_{\textrm{dd},is,\textrm{VV}}^\textrm{(2)}$, with $\hat{H}_{is}^\textrm{(2)}$ replacing $\hat{H}_{sn}$ on the right-hand side. We approximate the energy denominators with their spin-free asymptotic values, so that the whole of each operator is proportional to $R^{-3}$.

\begin{figure}[btp]
   \includegraphics[width=0.45\textwidth]{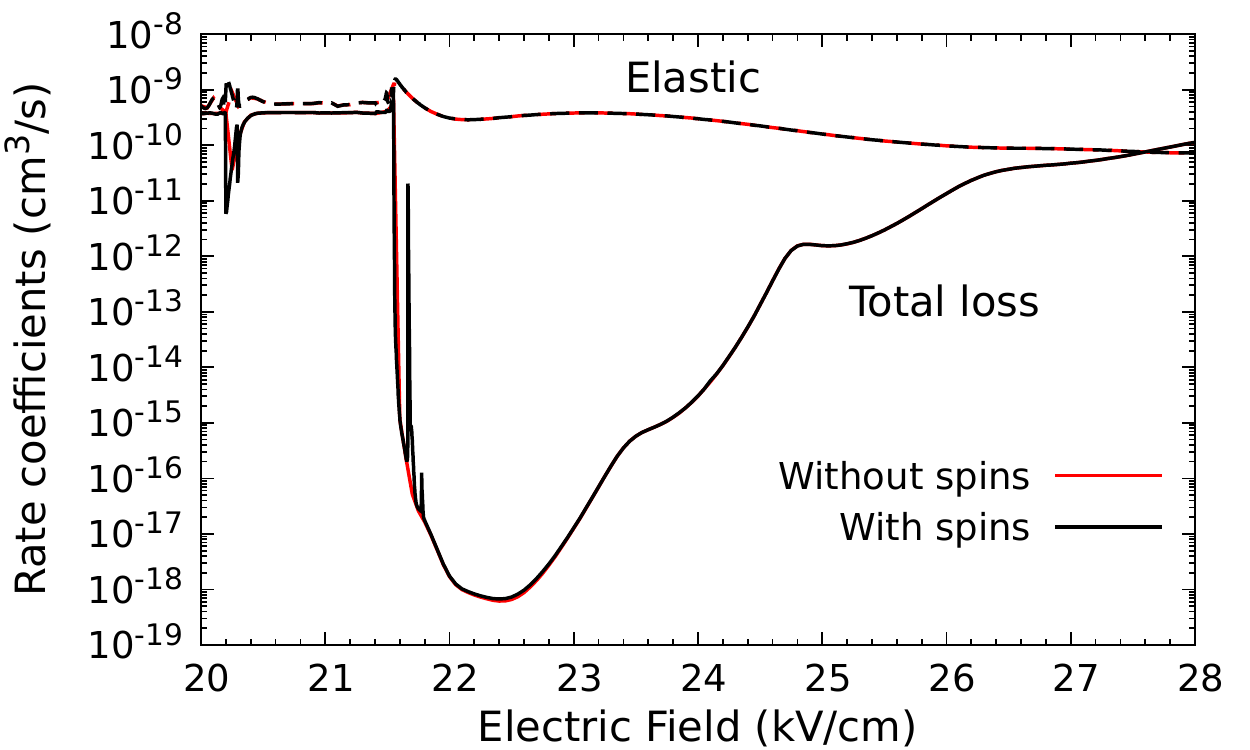}
    \caption{Effect of electron and nuclear spins on rate coefficients for CaF elastic collisions (dashed lines) and total loss (solid lines) for collision energy $E_\textrm{coll}/k_\textrm{B}=10\ \mu$K. The black curves use the large rotor basis set with $L_\textrm{max}=2$. Only $M_\textrm{tot}=0$ is included. The full spin structure is included for the pair functions in class 1, with the remainder included by Van Vleck transformations. The red curves show the corresponding results without spin structure, also for $L_\textrm{max}=2$.
}
\label{fig:rate_coeff-spins}
\end{figure}

Figure \ref{fig:rate_coeff-spins} shows rate coefficients for elastic scattering and total loss, with and without the inclusion of spins. These calculations use the large rotor basis set, but with $L_\textrm{max}$ restricted to 2 to make the calculations including spin affordable. All spin functions are included for every pair function in class 1. There is very little difference between the two calculations, \emph{except} in narrow regions of field around 21.7 and 21.8 kV/cm, where there are narrow spikes in the loss cross sections when spin is included. The similarity of the cross sections across the rest of the range of fields contrasts with the results shown for RbSr in Fig.\ 4 of Ref.\ \cite{Quemener:2016}, which showed substantially enhanced shielding at fields analogous to the upper half of Fig.\ \ref{fig:rate_coeff-spins}.

The large spin-induced loss peaks occur near the fields where spin-changing channels corresponding to (0,0)+(2,0) become energetically open. The large and small peaks correspond to open channels with one or both molecules, respectively, excited to states with $g=1$; the corresponding crossings are labeled with closed and open circles in Fig.\ \ref{fig:crossings}. There is a large flux into these outgoing channels when the kinetic energy of the products is very low (less than about 3~mK), and also at slightly lower fields, due to resonance effects described below. This is a dramatic, though localized, breakdown of the idea that spins act as spectators in the region important for shielding.

There are no matrix elements of spin operators that connect the incoming channels for (1,0)+(1,0) directly to the near-degenerate spin-changing channels. However, there \emph{are} second-order matrix elements of the type (\ref{eq:VV-spin}) that can cause such couplings, via other rotor states.
The operators $\hat{H}_{sn}$ and $\hat{H}_{is}^\textrm{(2)}$ act on the state of one monomer at a time and cannot change the quantum numbers of both monomers simultaneously. They are also diagonal in $L$ and $M_L$. $\hat{H}_{sn}$ can change $g$ and $m_g$ by 0 or $\pm1$ while conserving $m_n+m_g$, but does not have matrix elements diagonal in $m_n=0$ or $m_g=0$. It thus connects either (1,0)+(1,0) or (0,0)+(2,0) only to pair functions with $m_{n1}=\pm1$ or $m_{n2}=\pm1$. The only states of this type with lower energy are (0,0)+(1,$\pm1$), (0,0)+(2,$\pm1$), and (1,$\pm1$)+(1,0), but there are higher states too. $\hat{H}_{is}^\textrm{(2)}$ can change $m_g$ by 0, 1, or 2 while conserving $m_n+m_g$, but has no matrix elements involving $g=0$ in CaF. The overall effect is that $\hat{H}_{\textrm{dd},is,\textrm{VV}}^\textrm{(2)}$ has no matrix elements at all that connect directly to the initial state considered here, with $g_1=g_2=0$.

The principal approximation in our Van Vleck transformation is the approximation of the energy denominators in Eq.\ (\ref{eq:VV-spin}) by their asymptotic values. This can be important for nearby channels that come close in energy as a function of $R$. Such channels need to be in class 1 to capture their full effects. A minimal set of functions in class 1 to calculate the effects of spin is therefore $(\tilde{n},m_n,g,m_g)$ = (1,0,0,0)+(1,0,0,0) and the 12 channels obtained by combining (0,0,0,0) with (2,0,$g$,$m_g$) and (2,$\pm1$,$g$,$m_g$), with $(g,m_g)$ taking all 4 possible values. We refer to this basis set as spin-N13.

\begin{figure}[tbp]
   \includegraphics[width=0.5\textwidth]{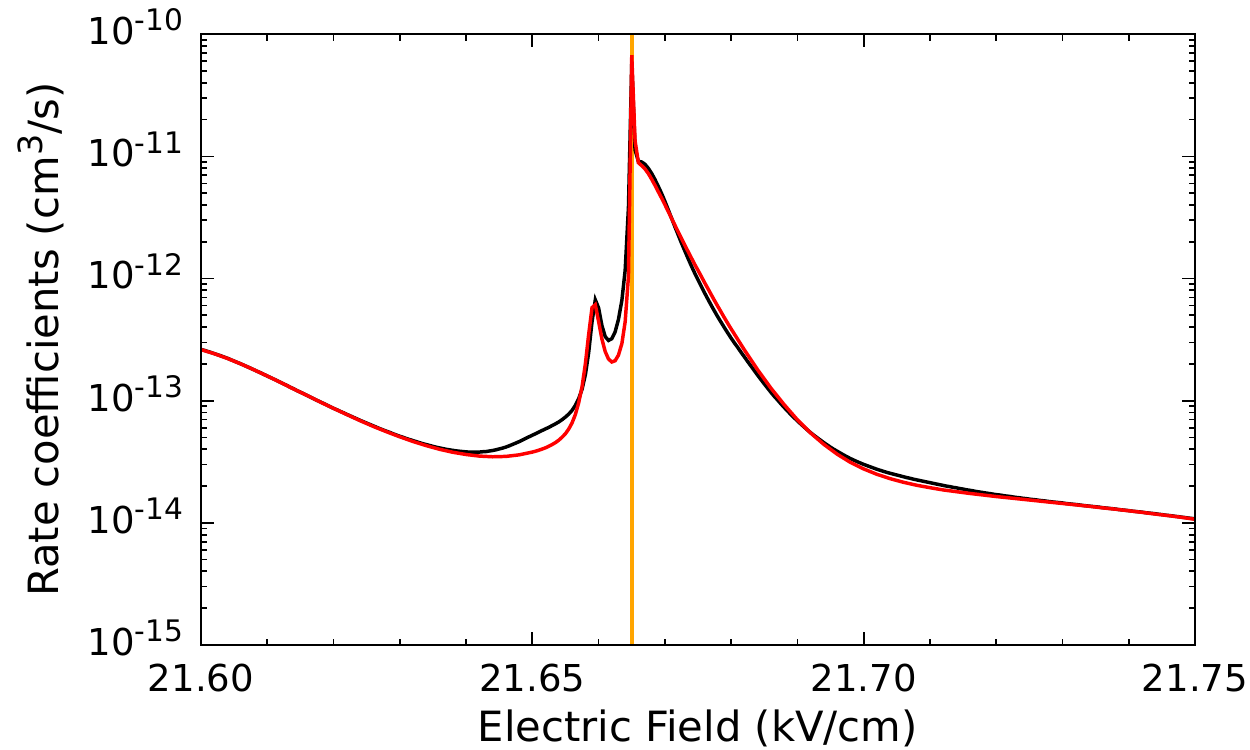}
    \caption{Rate coefficient for total loss at $E_\textrm{coll}/k_\textrm{B}=10\ \mu$K, calculated with the small rotor basis set combined with full spin structure (black) and with the spin-N13 basis set (red).  The calculations use $L_\textrm{max}=20$ and $M_\textrm{tot}=0$. The vertical orange line shows the field where the spin-changing channels open.
}%
\label{fig:spins-cs-vv}
\end{figure}

Figure \ref{fig:spins-cs-vv} shows the spin-induced loss peak near 21.66 kV/cm in more detail. It compares results using the spin-N13 basis set (red line) with those obtained using the small rotor basis set combined with all possible spin functions in class 1 (black line). All other channels up to $\tilde{n}=5$ are included via Van Vleck transformations at minimal extra cost. These calculations use $L_\textrm{max}=20$, which gives much better convergence than in Fig.\ \ref{fig:rate_coeff-spins}, with a larger background loss. The vertical orange line shows the field where the spin-changing channels corresponding to (0,0,0,0)+(2,0,1,$\pm1$) become energetically open.
The lowest such channels for $M_\textrm{tot}=0$ have $L=2$, with centrifugal barriers of height 13 $\mu\textrm{K} \times k_\textrm{B}$ near 900~$a_0$. The rate coefficient shows a sharp peak when the kinetic energy release is near this barrier maximum, then drops smoothly back to the background rate over the next 0.1 kV/cm. There is also a peak near 21.66 kV/cm, where the spin-changing channels are still closed; this is a Feshbach resonance due to a state bound by about 300 $\mu\textrm{K} \times k_\textrm{B}$ in each spin-changing channel.
The spin-N13 basis set successfully reproduces the full pattern of peaks, including the resonance. Calculations with the spin-N13 basis set take about a factor of 100 less computer time than those with the full spin basis set, so this shows an impressive further demonstration of the power of the Van Vleck transformation.

\begin{figure}[tbp]
   \includegraphics[width=0.5\textwidth]{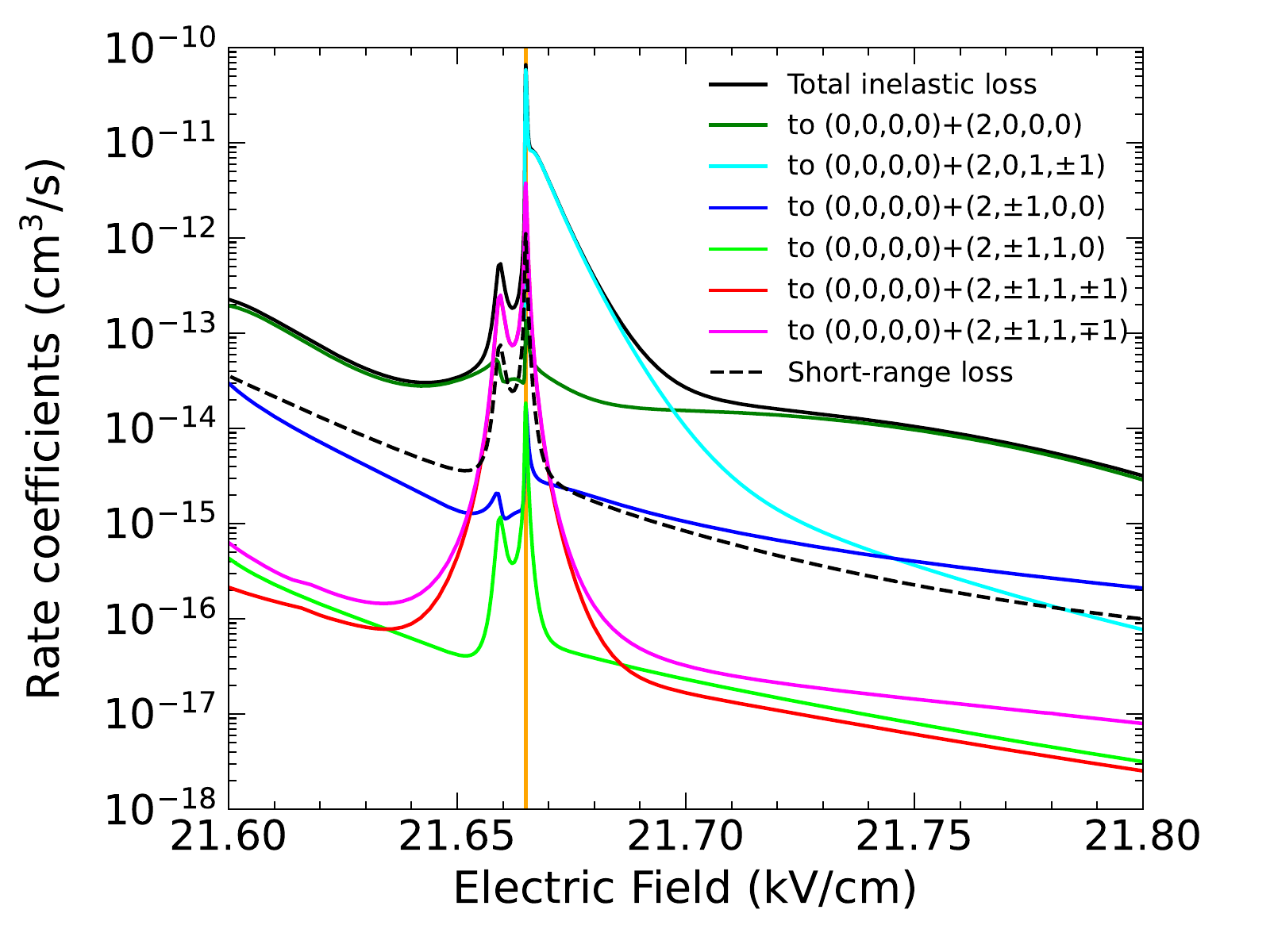}
    \caption{Rate coefficients at $E_\textrm{coll}/k_\textrm{B} = 10\ \mu$K around the spin-changing loss peak, obtained with the spin-N13 basis set. The black solid line shows the total inelastic rate coefficient and the colored lines show the state-to-state inelastic rate coefficients from the initial state (1,0,0,0)+(1,0,0,0) to other levels. The dashed black line shows the rate coefficient for short-range loss. The vertical orange line shows the field where the spin-changing channels open.
    }
\label{fig:spins-st-to-st-xs}
\end{figure}

Figure \ref{fig:spins-st-to-st-xs} shows rate coefficients for state-to-state inelastic processes and short-range loss around the spin-induced peak. Far from the peak on both sides, the dominant product states are (0,0,0,0)+(2,0,0,0) and (0,0,0,0)+(2,$\pm1$,0,0), which are driven directly by spin-free dipole-dipole interactions. The vertical line shows the field at which the spin-changing channels open. Immediately above this field, the total cross section is dominated by flux into the newly open channels. Just below this field, the products are mostly in lower states (0,0,0,0)+(2,$m_n$,1,$\pm1$). These pair states are directly coupled to the resonant channels corresponding to (0,0,0,0)+(2,0,1,$\pm1$) by $\hat{H}_\textrm{dd}$, so flux into them is enhanced when there is additional density in (0,0,0,0)+(2,0,1,$\pm1$) close to the resonances. There is very little flux into (0,0,0,0)+(2,0,1,0), although it becomes open in the same range of fields.

The positions of the spin-changing peaks depend strongly on the molecular coupling constants, particularly the hyperfine splitting. For CaF this splitting is small, around 120 MHz, so the main spin-changing peak is only about 0.11 kV/cm above the crossing field. SrF is similar in this respect. For some other $^2\Sigma$ molecules, however, the hyperfine splitting is considerably larger \cite{Aldegunde:doublet:2018} and there may be spin-changing peaks that lie at fields that will interfere with shielding.

\begin{figure}[tbp]
   \includegraphics[width=0.5\textwidth]{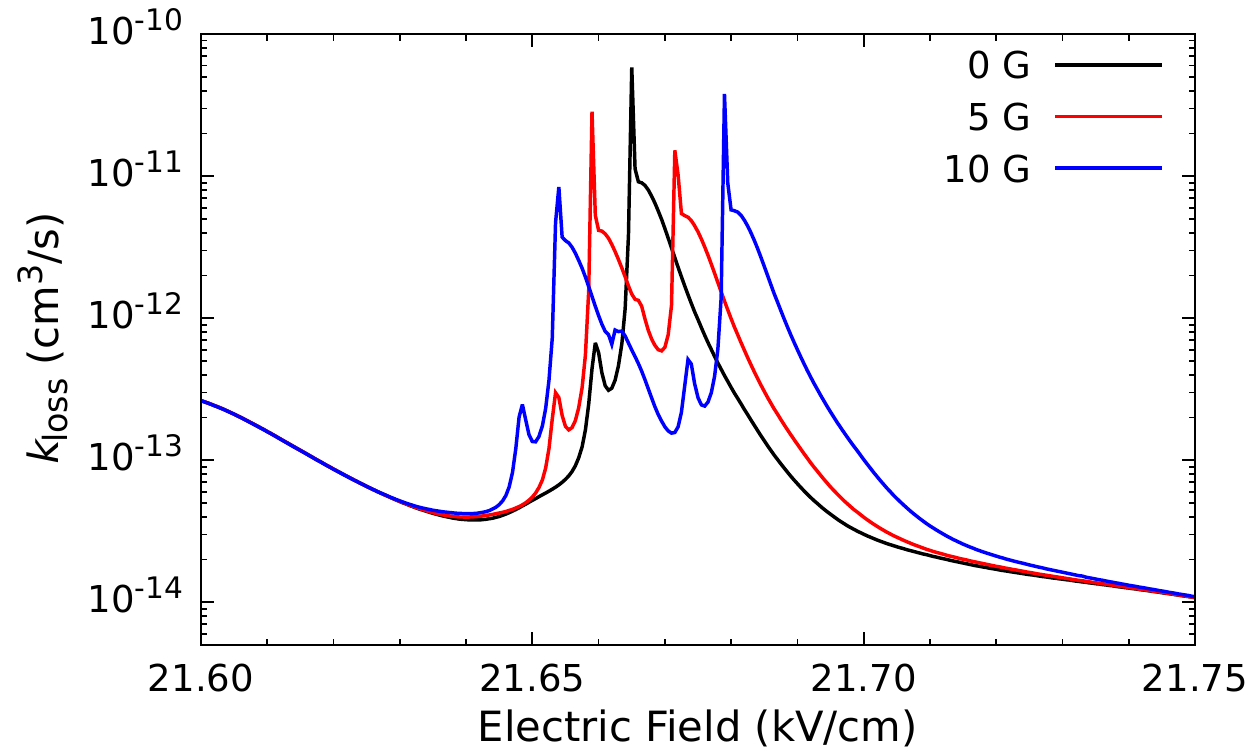}
    \caption{Effect of a small magnetic field on the rate coefficient for total loss at $E_\textrm{coll}/k_\textrm{B} = 10\ \mu$K. The calculations use the small rotor basis set with full spin structure, with $L_\textrm{max}=20$ and $M_\textrm{tot}=0$. The loss rate is almost unaffected by magnetic fields outside the range shown.
}%
\label{fig:magneticfld}
\end{figure}

\subsection{Effect of magnetic field}
\label{sec:fields}

It is important to know whether magnetic fields have significant effects on shielding. In the case of microwave shielding of CaF, a magnetic field around 100~G is beneficial because it recovers a nearly conserved quantum number $m_n$ that is otherwise destroyed by $\hat{h}_\textrm{fhf}$. For static shielding, this is not an issue because the electric field itself is sufficient to ensure that $m_n$ is nearly conserved. Nevertheless, small magnetic fields may cause appreciable splitting of otherwise near-degenerate levels, and we investigate those effects here.

A magnetic field $\boldsymbol{B}$ introduces Zeeman terms in the Hamiltonian. In a full treatment \cite{Caldwell:long-coh:2020}, there are terms involving the electron spin, the nuclear spin, and the molecular rotation. However, the rotational and nuclear-spin terms are typically three orders of magnitude smaller than the electron-spin term. We therefore ignore them in the present calculations, and consider only the term
\begin{align}
\hat{h}_{\mathrm{Zeeman}} =
g_S\mu_\mathrm{B} \boldsymbol{\hat{s}} \cdot \boldsymbol{B}.
\end{align}
A magnetic field has very little effect on the rate coefficients, except near the peak involving low-energy spin-changing channels described in Sec. \ref{sec:spins}. However, it does substantially modify this peak. Figure \ref{fig:magneticfld} shows the rate coefficients for total loss in this region in magnetic fields $B=5$ and 10~G, parallel to the electric field, with those in zero magnetic field. It may be seen that the peak splits into two similar structures, with a separation approximately proportional to the field. This occurs because the state (0,0,0,0)+(2,0,1,$\pm1$), which is responsible for the feature as described in Sec. \ref{sec:spins}, it itself split into two Zeeman components with $m_g=+1$ and $-1$.

\section{Conclusions}
\label{sec:conc}

We have studied ultracold collisions of two CaF molecules in high static electric fields. A near degeneracy between field-dressed pair functions allows the creation of a repulsive long-range barrier due to dipole-dipole forces. This barrier prevents the molecules reaching the short-range region where inelastic and other loss processes are likely to occur.

We have carried out coupled-channel quantum scattering calculations on the ultracold collisions. Electric fields cause strong mixing of CaF rotor states, so we use basis sets based on field-dressed rotor states. Converged calculations require large basis sets of both rotor states and partial waves, and can quickly become prohibitively expensive in computer time. We have developed an efficient way to include energetically well-separated rotor states in the calculation, using a Van Vleck transformation, so that their effects are taken into account without including extra functions in the coupled-channel basis set. With this method, calculations with large basis sets are made much cheaper and even very small explicit basis sets can give results of useful accuracy.

We have found that static-field shielding is particularly effective for CaF. Loss processes are reduced by up to seven orders of magnitude over a broad range of collision energies. The rate coefficients for loss reach a minimum near 23 kV/cm, and are suitable for efficient evaporative cooling all the way to Bose-Einstein condensation. At the lowest collision energies, the rate coefficients remain below 10$^{-13}$ cm$^3$\,s$^{-1}$ across a range of fields from 21.7 to 24.5 kV/cm. This should allow production of long-lived dipolar condensates with space-fixed molecular dipole moments tunable from $-0.44$ to $-0.34$~D.

We have studied the energy dependence of different contributions to rate coefficients for elastic scattering and loss. The elastic rate at the lowest energies shows a pronounced peak near 23 kV/cm, principally because of a maximum in the s-wave scattering length there. At collision energies close to 10 $\mu$K, by contrast, the s-wave contribution is small; elastic scattering is then dominated by higher partial waves and the dependence on field is much weaker. Loss processes are mostly dominated by s-wave scattering. The loss rate is almost independent of energy below 100~nK, but it decreases substantially at higher energies; at 23 kV/cm it decreases by about a factor of 20 between 10~nK and 10~$\mu$K.

We have investigated the effects of electron and nuclear spin on shielding collisions. At most fields the effects are very small. However, there are dramatic enhancements of loss rates near specific fields where spin-changing channels are energetically just accessible. At such fields the spins are intimately involved in the collision, and are far from being ``spectators". For CaF, such a feature exists at fields just below 21.7 kV/cm, where it enhances loss rates by up to three orders of magnitude. We have also investigated the effects of small magnetic fields, which modify the spin-changing loss feature but otherwise have little effect on rate coefficients.

This paper paves the way for experiments on evaporative cooling of CaF in strong electric fields. It shows that such experiments have a good prospect of cooling CaF all the way to quantum degeneracy. While the BEC will be stable against two-body loss processes for several seconds, it may not be stable against collapse. The stability and behavior of the condensate depend on the dipole length $D$ and the s-wave scattering length $a$.  We calculate a negative scattering length for all electric fields and collision energies of interest. In contrast with unshielded systems, where $a$ is determined by short-range physics and cannot be predicted from \emph{a priori} potentials, $a$ is here determined by the long-range well that exists outside the shielding barrier. Because of this, we expect the present calculation of $a$ to be accurate.

In free space, and in the absence of the electric shield, a dipolar BEC will collapse when the scattering length is negative, because the energy is lowered as the density increases indefinitely. The collapse can be avoided by confining the BEC in a pancake-shaped trap with the dipoles aligned along the short axis, so that most dipoles lie side-by-side and repel \cite{Santos:2000, Koch:2008}. The electric shield prevents close approach of two molecules, so may also help to stabilize the BEC against collapse, although its influence on the many-body dynamics of a BEC has not yet been studied.  Further stabilization can come from fluctuations around the mean-field energy, and in some circumstances a dipolar BEC can form self-bound droplets and exotic supersolid phases \cite{Schmidt:2022}. The stability and many-body phases of a strongly dipolar BEC in the presence of an electric shield and in various trap geometries are interesting topics for future study.

\section*{Rights retention statement}

For the purpose of open access, the authors have applied a Creative Commons Attribution (CC BY) licence to any Author Accepted Manuscript version arising from this submission.

\section*{Data availability statement}

The data presented in this paper are available from Durham University~\cite{DOI_data-CaF-shielding}.

\section*{Acknowledgement}
We are grateful to Dr German Sinuco Leon for preliminary calculations on this system.
This work was supported by the U.K. Engineering and Physical Sciences Research Council (EPSRC) Grant Nos.\ EP/P01058X/1, EP/W00299X/1, EP/V011499/1 and EP/V011677/1.

%\appendix

\section*{Appendix: Convergence of scattering calculations}

\subsection{Calculation of cross sections}
\label{sec:x-sec}

Elastic and state-to-state inelastic cross sections involving initial and final pair levels $i=(\tilde{n}_1,m_{n1},\tilde{n}_2,m_{n2})$ and $f=(\tilde{n}'_1,m'_{n1},\tilde{n}'_2,m'_{n2})$ are obtained in terms of S-matrix elements from coupled-channel calculations,
\begin{align}
\sigma_{\textrm{el},i} &= \frac{g\pi}{k_0^2} \sum_{LL'M_LM'_LM_\textrm{tot}} \left| \delta_{LL'} \delta_{M_LM'_L} - S^{M_\textrm{tot}}_{iLM_L,i L'M'_L}\right|^2; \\
\sigma_{\textrm{inel},if} &= \frac{g\pi}{k_0^2} \sum_{LL'M_LM'_LM_\textrm{tot}} \left|  S^{M_\textrm{tot}}_{iLM_L,fL'M'_L}\right|^2,
\end{align}
where $g=2$ for identical bosons and $k_0=(2\mu E_\textrm{coll}/\hbar^2)^{1/2}$ is the incoming wave vector. Where necessary, spin quantum numbers are included in the specification of states $i$ and $f$. Total inelastic cross sections $\sigma_{\textrm{inel},i}$ are obtained by summing over all final pair levels $f\ne i$ at long range.

The cross section for short-range loss is obtained from the unitarity deficit, summed over channels for incoming state $i$,
\begin{equation}
\sigma_{\textrm{short},i} = \frac{g\pi}{k_0^2} \sum_{LM_LM_\textrm{tot}} \left( 1 - \sum_{fL'M'_L} \left| S^{M_\textrm{tot}}_{iLM_L,fL'M'_L} \right|^2 \right),
\label{eq:unit-def}
\end{equation}
where here the sum over $f$ includes $i$. The short-range loss may include contributions from inelastic processes that occur inside $R_\textrm{absorb}$. The total loss may be calculated either as the sum of $\sigma_{\textrm{inel},i}$ and $\sigma_{\textrm{short},i}$ or equivalently as
\begin{equation}
\sigma_{\textrm{loss},i} = \frac{g\pi}{k_0^2} \sum_{LM_LM_\textrm{tot}} \left( 1 - \sum_{L'M'_L} \left| S^{M_\textrm{tot}}_{iLM_L,iL'M'_L} \right|^2 \right).
\label{eq:xs-total}
\end{equation}

Partial cross sections for a single incoming $L$, designated $L_\textrm{in}$ elsewhere for clarity, are obtained from similar expressions without the sum over $L$.

\subsection{Convergence with respect to $L_\textrm{max}$}
\label{sec:Lmax}

\begin{figure}[tbp]
\centering
\includegraphics[width=0.45\textwidth]{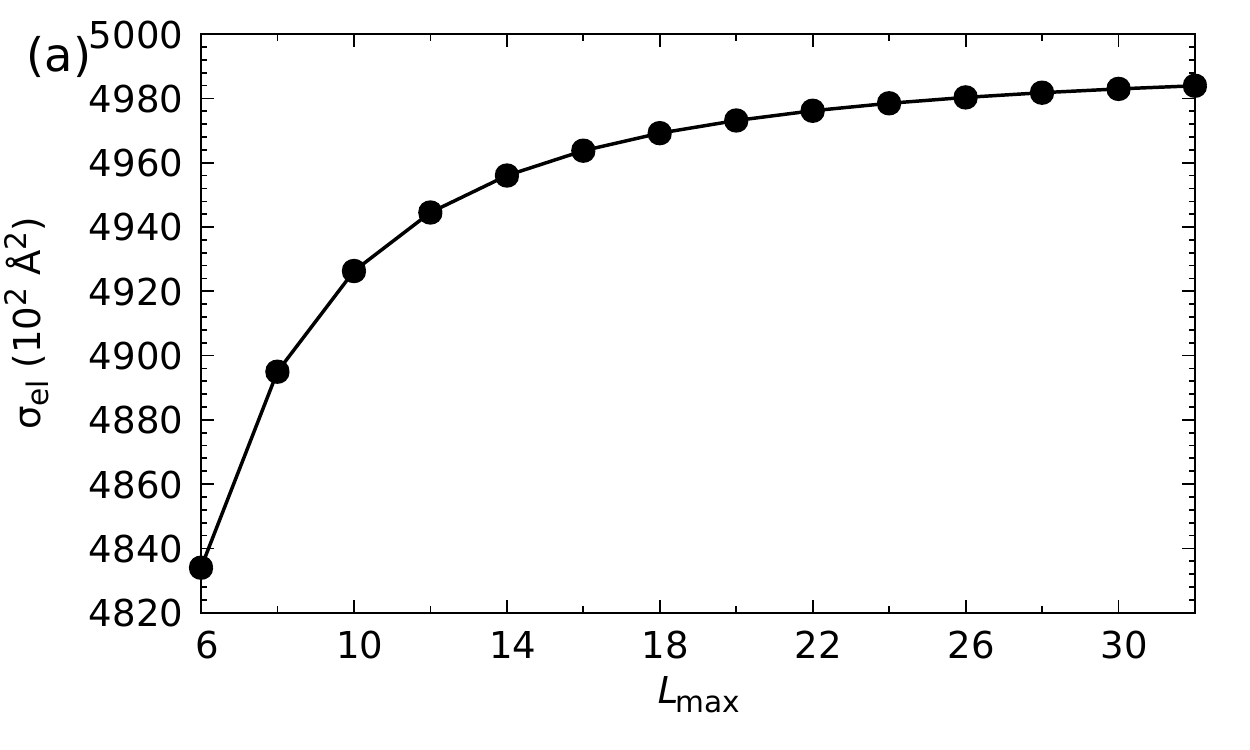}
\includegraphics[width=0.45\textwidth]{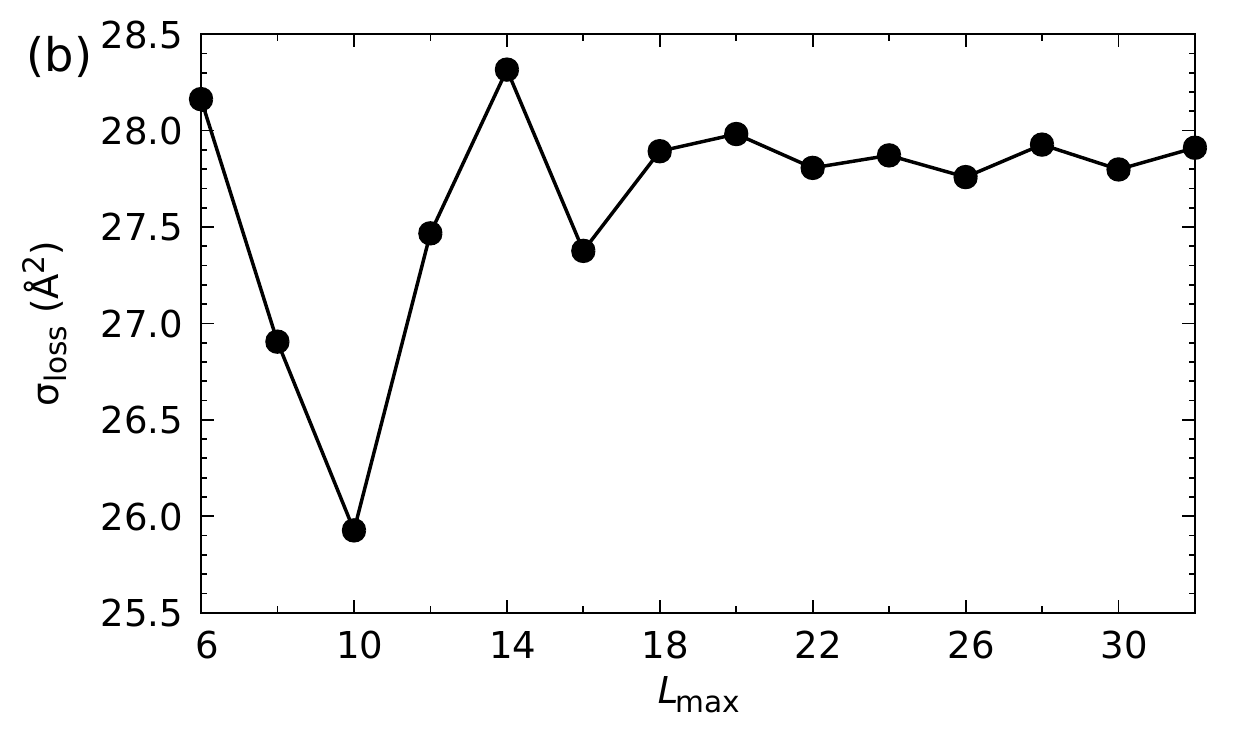}
\caption{Convergence of cross sections for (a) elastic scattering and (b) total loss with respect to $L_\textrm{max}$ at electric field $F=24.5$~kV/cm. The calculations use the large rotor basis set.
}
\label{fig:Lmax}
\end{figure}

\begin{figure}[tbp]
\centering
\includegraphics[width=0.45\textwidth]{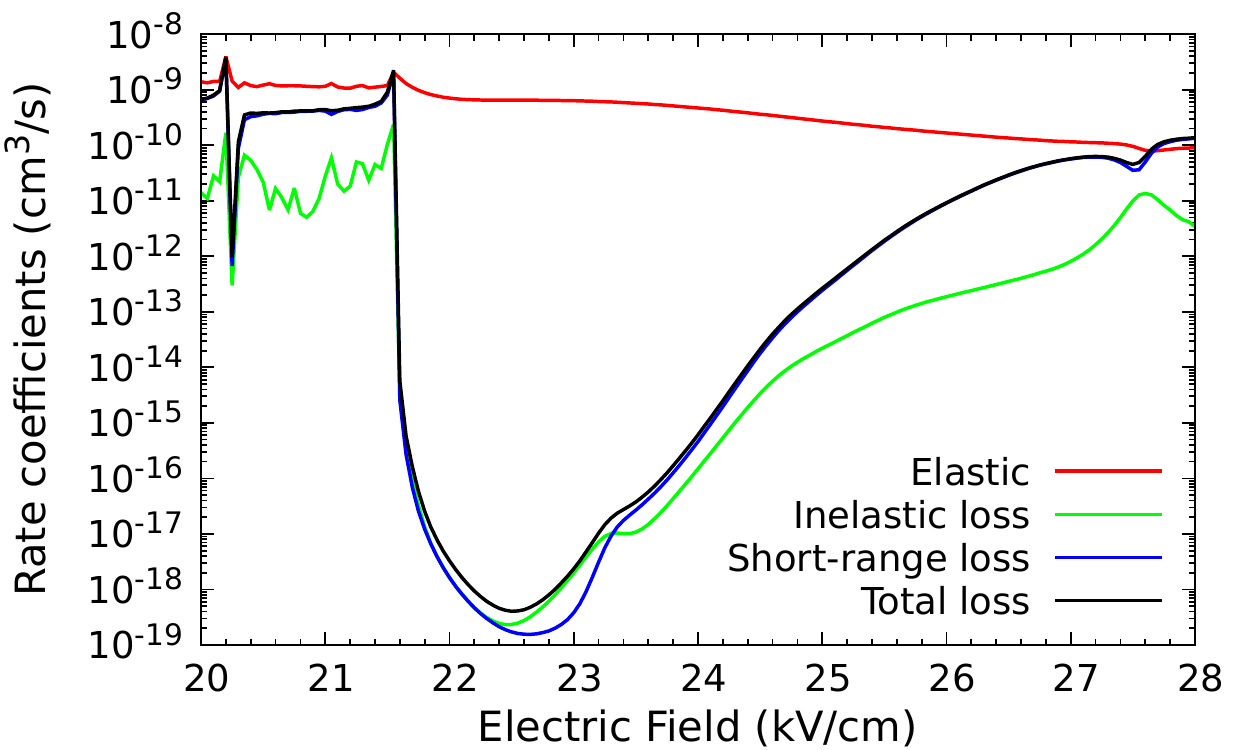}
\caption{Rate coefficients for spin-free CaF elastic collisions and loss processes as a function of electric field for $E_\textrm{coll}/k_\textrm{B}=10\ \mu$K. The calculations use the large rotor basis set with $L_\textrm{max}=6$.
}
\label{fig:rate_coeff6}
\end{figure}

Figure \ref{fig:Lmax} shows the convergence of the cross sections for elastic scattering and total loss with respect to $L_\textrm{max}$ at 10 $\mu$K and 24.5 kV/cm. The elastic cross sections $\sigma_\textrm{el}$ converge quite fast with respect to $L_\textrm{max}$, and are converged to within 1\% of their final value by $L_\textrm{max}=12$. However, the loss cross sections converge much more slowly, and require $L_\textrm{max}=18$ for a similar degree of convergence. In most calculations we use $L_\textrm{max}=20$.

Figure \ref{fig:rate_coeff6} shows rate coefficients for collision energy $E_\textrm{coll} = 10\,\mu\textrm{K} \times k_\textrm{B}$, calculated with the same basis set of rotor functions as Fig.\ \ref{fig:rate_coeff}, but with $L_\textrm{max}=6$ instead of $L_\textrm{max}=20$. It gives qualitatively correct results, but underestimates the rate coefficients for short-range and inelastic loss by about an order of magnitude around 23 kV/cm, where maximum shielding occurs.

\subsection{Convergence with respect to basis set of rotor functions}
\label{sec:rotor-basis}

\begin{figure*}[tbp]
	\subfloat[]{
		\includegraphics[width=0.45\textwidth]{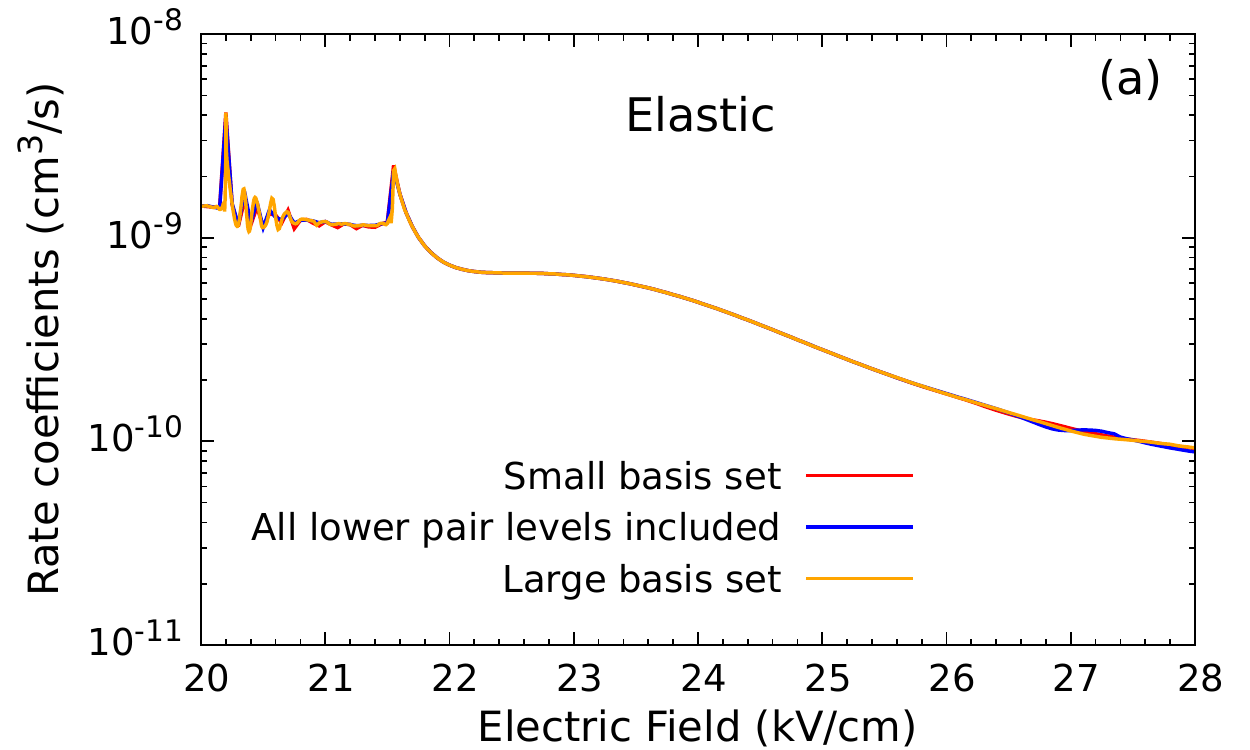}
	}
%    \vspace{-0.8 cm}
	\subfloat[]{
		\includegraphics[width=0.45\textwidth]{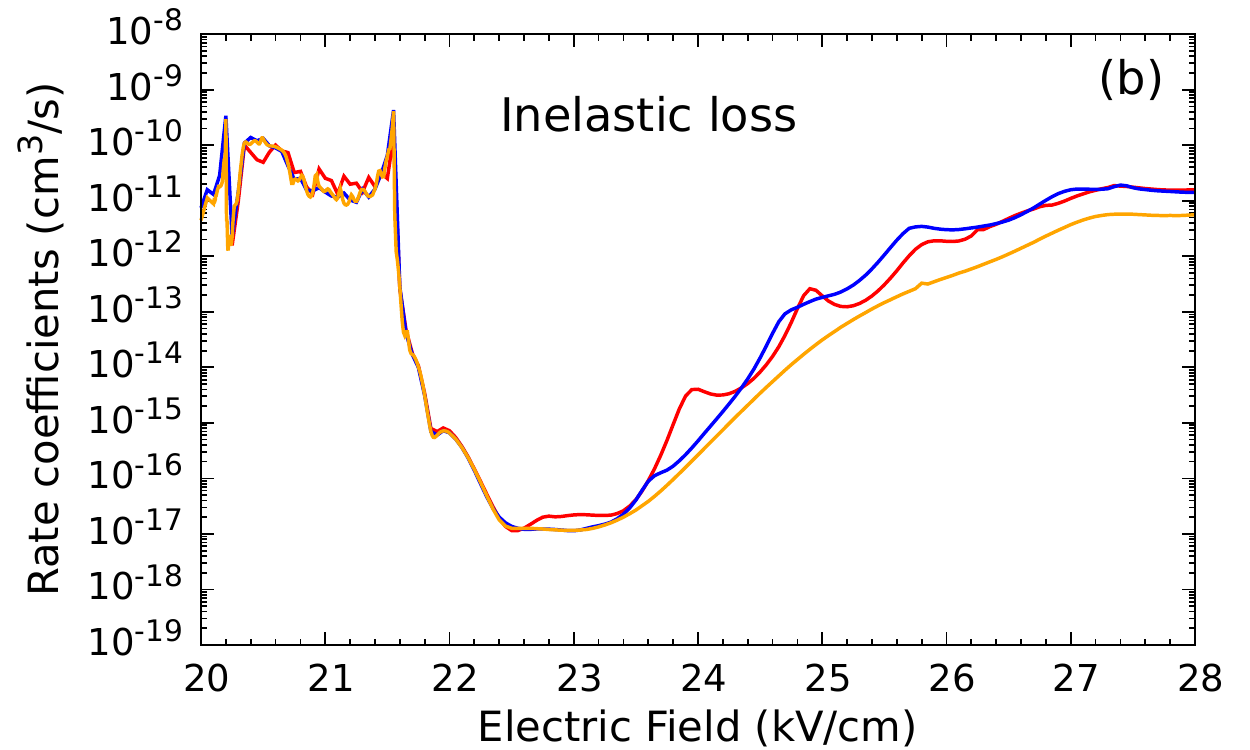}
	}
    \vspace{-0.8 cm}
	\subfloat[]{
		\includegraphics[width=0.45\textwidth]{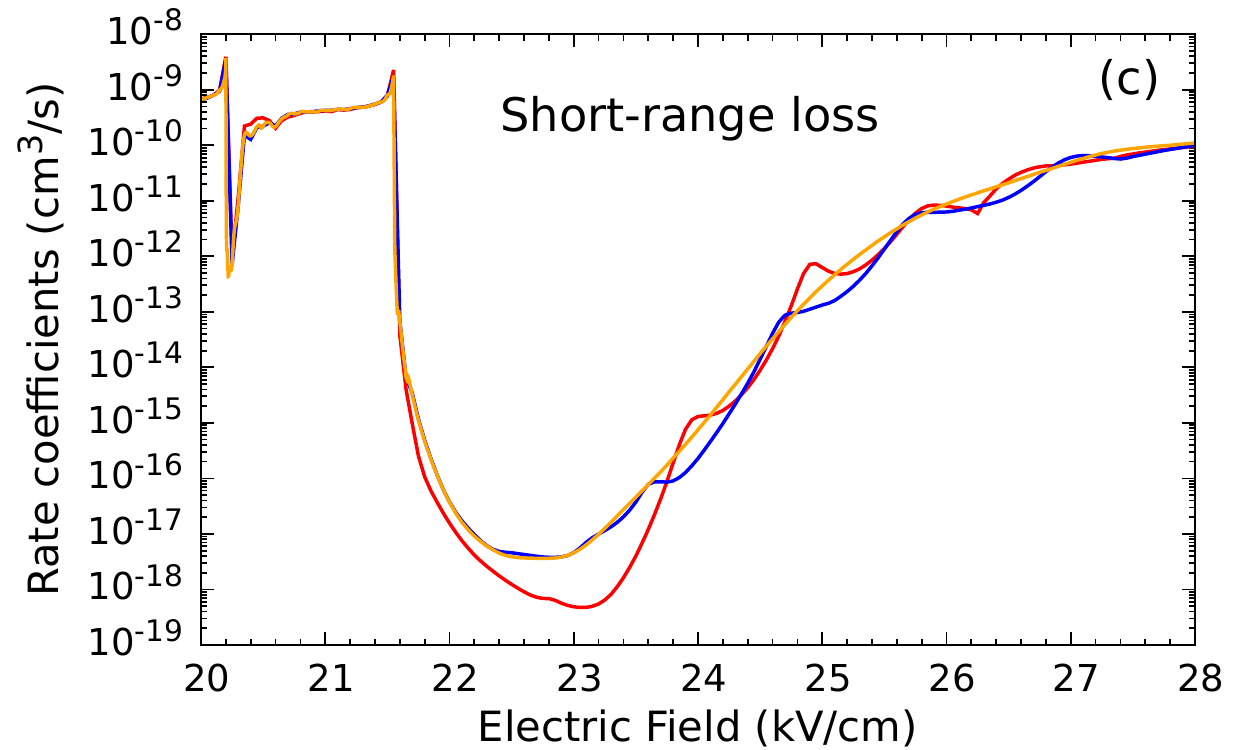}
	}
%    \vspace{-0.8 cm}
	\subfloat[]{
		\includegraphics[width=0.45\textwidth]{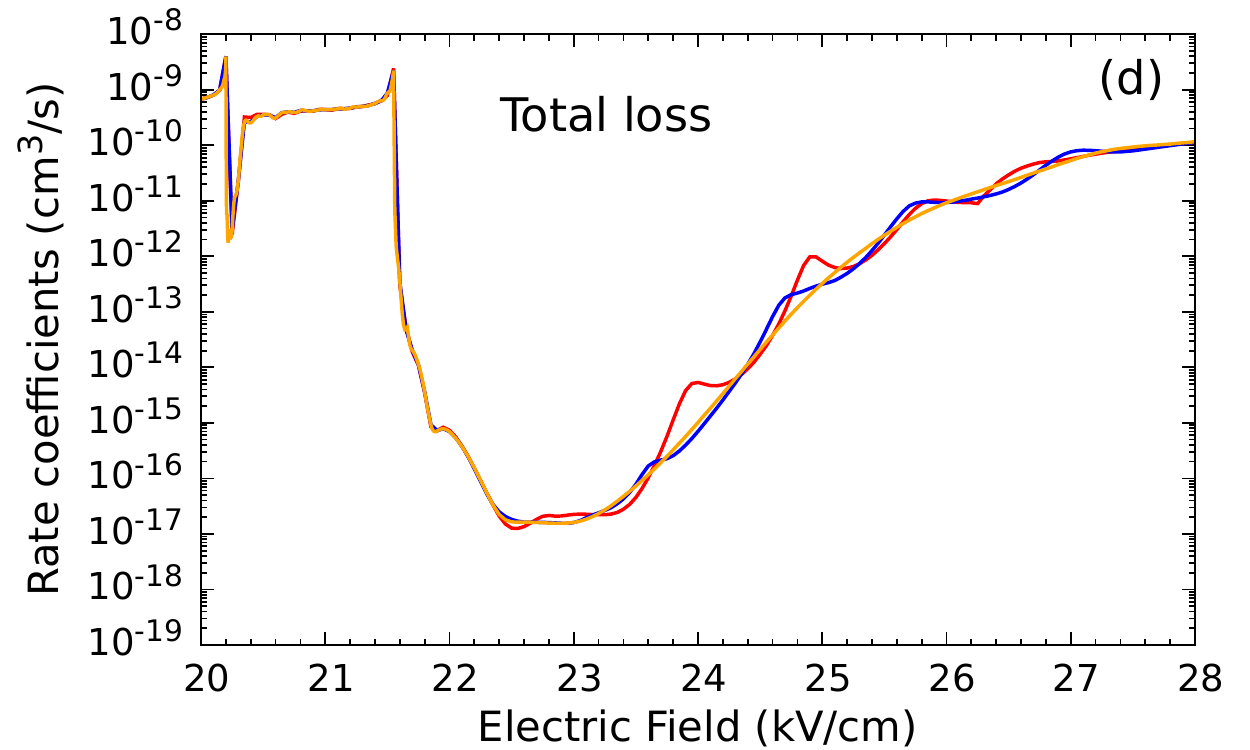}
	}
%    \vspace{-0.4 cm}
\caption{Dependence of rate coefficients on the basis set of rotor functions included in class 1. All calculations use $\tilde{n}_\textrm{max}=n_\textrm{max}=5$ and $L_\textrm{max}=20$ and are at $E_\textrm{coll}/k_\textrm{B}=10~\mu$K. Yellow: large rotor basis set, with all pair levels up to $\tilde{n}=2$ included in class 1: this is the basis set used for most calculations in the present paper.
Blue: all pair levels below (1,0)+(1,0) and (0,0)+(2,0) in class 1.
Red: small rotor basis set, with only pair levels (1,0)+(1,0), (0,0)+(2,0), and (0,0)+(2,$\pm$1) in class 1.}
\label{fig:class-I-convergence}
\end{figure*}

Figure \ref{fig:class-I-convergence} shows the dependence of the coupled-channel rate coefficients on the set of pair functions included in class 1. All remaining basis functions up to $\tilde{n}_\textrm{max}=n_\textrm{max}=5$ are included in class 2 and are accounted for by the Van Vleck transformation. The orange curves show results with the large rotor basis set used for most calculations in the present paper, with all pair levels up to $\tilde{n}=2$ included in class 1. The blue curves show results with a smaller rotor basis set with all pair levels below (1,0)+(1,0) and (0,0)+(2,0) in class 1. The red curves show results with the small rotor basis set, with only (1,0)+(1,0), (0,0)+(2,0) and (0,0)+(2,$\pm1$) in class 1.

\begin{figure}[tbp]
\includegraphics[width=0.49\textwidth]{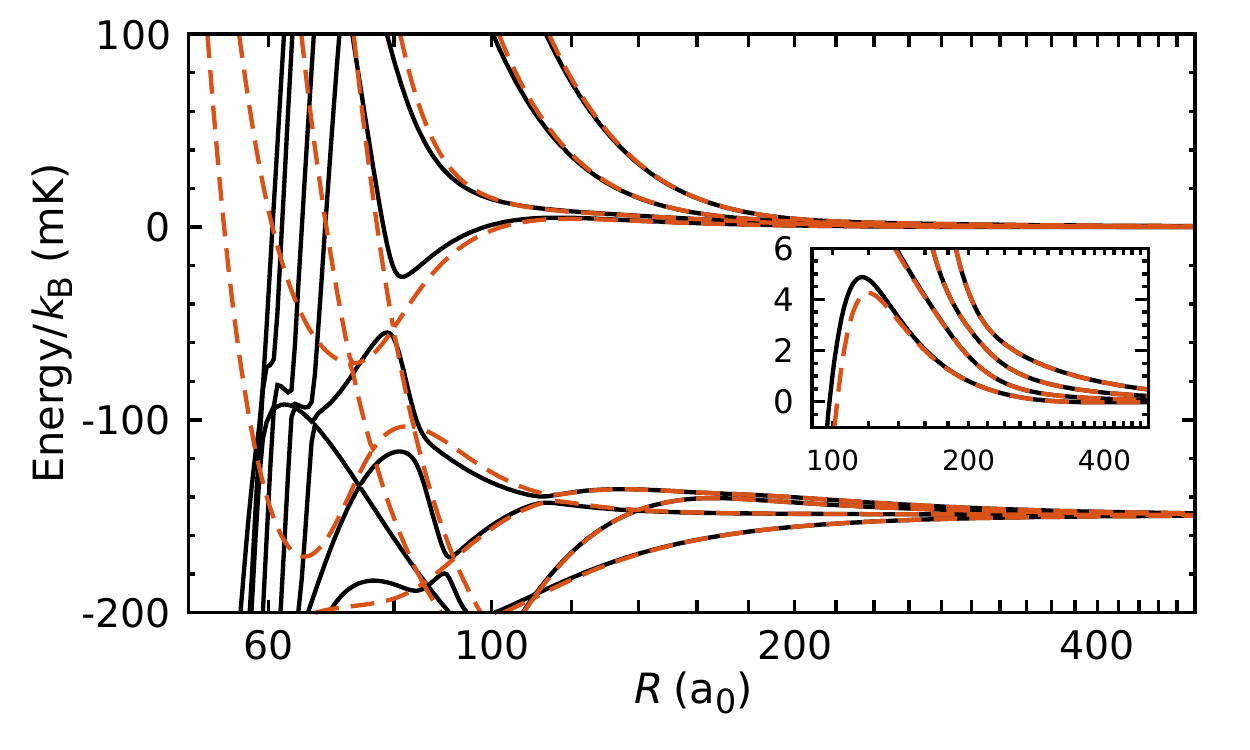}
\caption{Adiabats correlating with pair levels (1,0)+(1,0) and (0,0)+(2,0), calculated for an electric field of 24.5 kV/cm with $L_\textrm{max}=6$ by full diagonalization with $n_\textrm{max}=5$ (black solid lines) and with the small rotor basis set (orange dashed lines). The inset shows an expanded view of the adiabats near the long-range barrier for incoming $L=0$.
}
\label{fig:adiabats-minimal}
\end{figure}

Figure \ref{fig:class-I-convergence}(a) shows the elastic cross sections, and it may be seen that even the smallest rotor basis set gives good results for these. Figures.\ \ref{fig:class-I-convergence}(b), \ref{fig:class-I-convergence}(c), and \ref{fig:class-I-convergence}(d) show short-range loss, inelastic loss and total loss, respectively, with total loss being the most important. Even the small rotor basis set (red curves) gives qualitatively correct results, and is about a factor of 1500 cheaper than the large rotor basis set. However, it does introduce significant oscillations in the loss cross sections between 23 and 28 kV/cm. The oscillations are approximately in phase for inelastic scattering and short-range loss. They are resonant oscillations due to states confined inside the barrier in the adiabat for $L=0$. The adiabats for the small rotor basis set are compared with those for the large rotor basis set in Fig.\ \ref{fig:adiabats-minimal}. In both cases there is a classically allowed region inside the barrier, extending from at $R\sim 100\ a_0$ inwards. For the small rotor basis set (orange dashed curves) there is a simple potential well in this region, extending to $R\sim 60\ a_0$. Barrier penetration is enhanced near states confined in this well, and produces increases in both inelastic scattering and short-range loss. When extra rotor functions are added, however, they introduce additional avoided crossings between adiabats (black curves), which complicate the short-range reflections. As a result, the oscillations are only just visible for the large rotor basis set.

Even a minimal basis set, with only (1,0)+(1,0) and (0,0)+(2,0) in class 1 (not shown) gives qualitatively correct results, although in this case the oscillations between 23 and 27 kV/cm are even more pronounced and the loss rates deviate from the large-basis results by up to a factor of 10 at some fields.

\subsection{Behavior with respect to $R_\textrm{absorb}$}
\label{sec:Rmin}

\begin{figure*}[tbp]
	\subfloat[]{
		\includegraphics[width=0.45\textwidth]{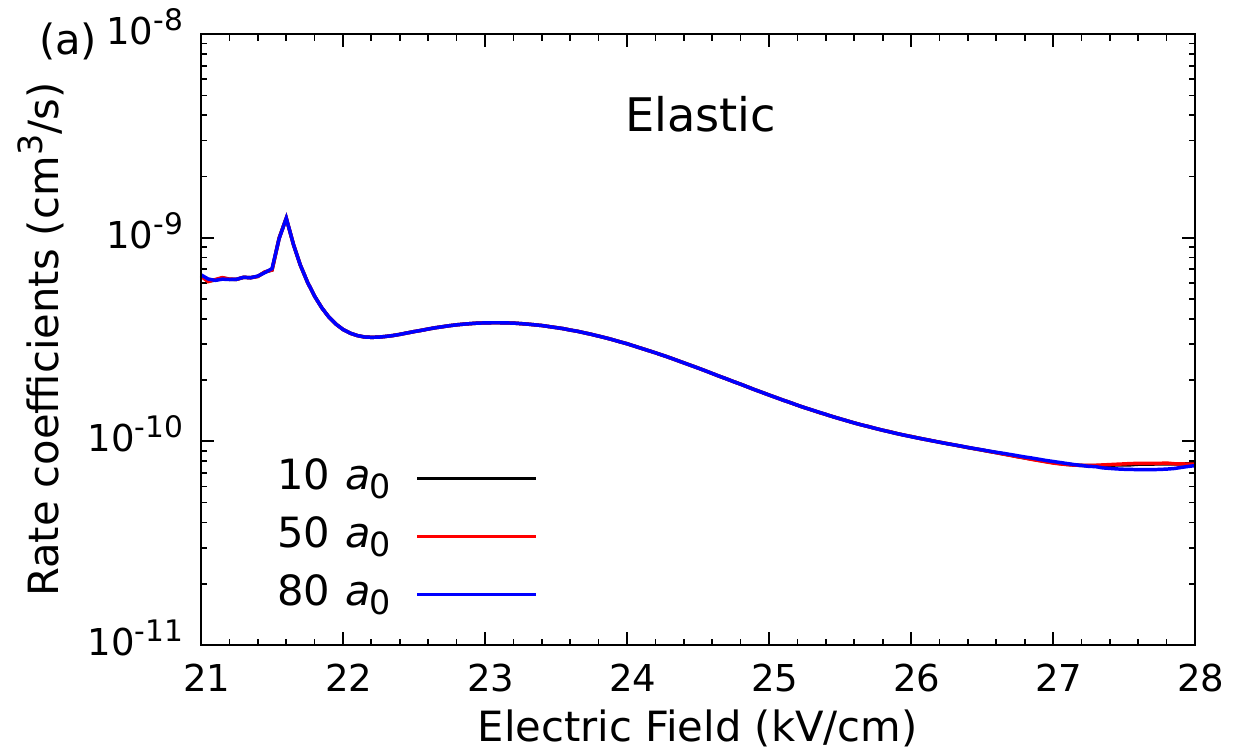}
	}
%    \vspace{-0.8 cm}
	\subfloat[]{
		\includegraphics[width=0.45\textwidth]{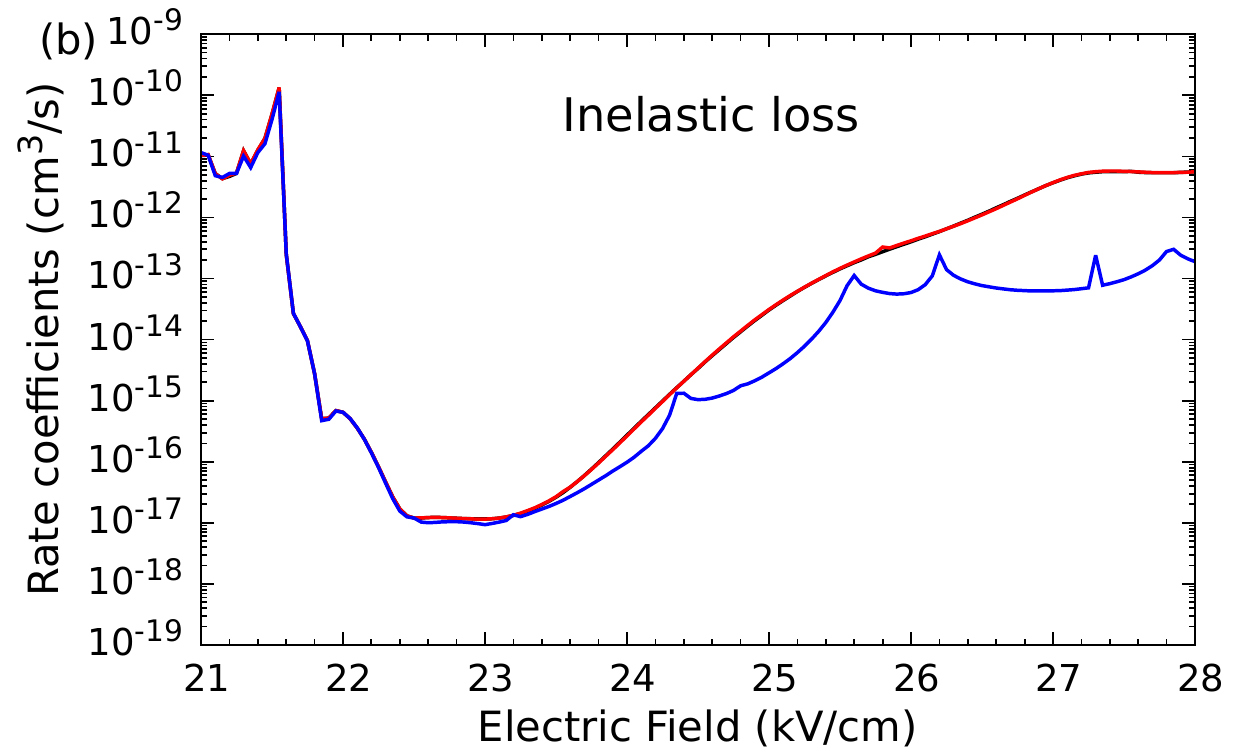}
	}
    \vspace{-0.8 cm}
	\subfloat[]{
		\includegraphics[width=0.45\textwidth]{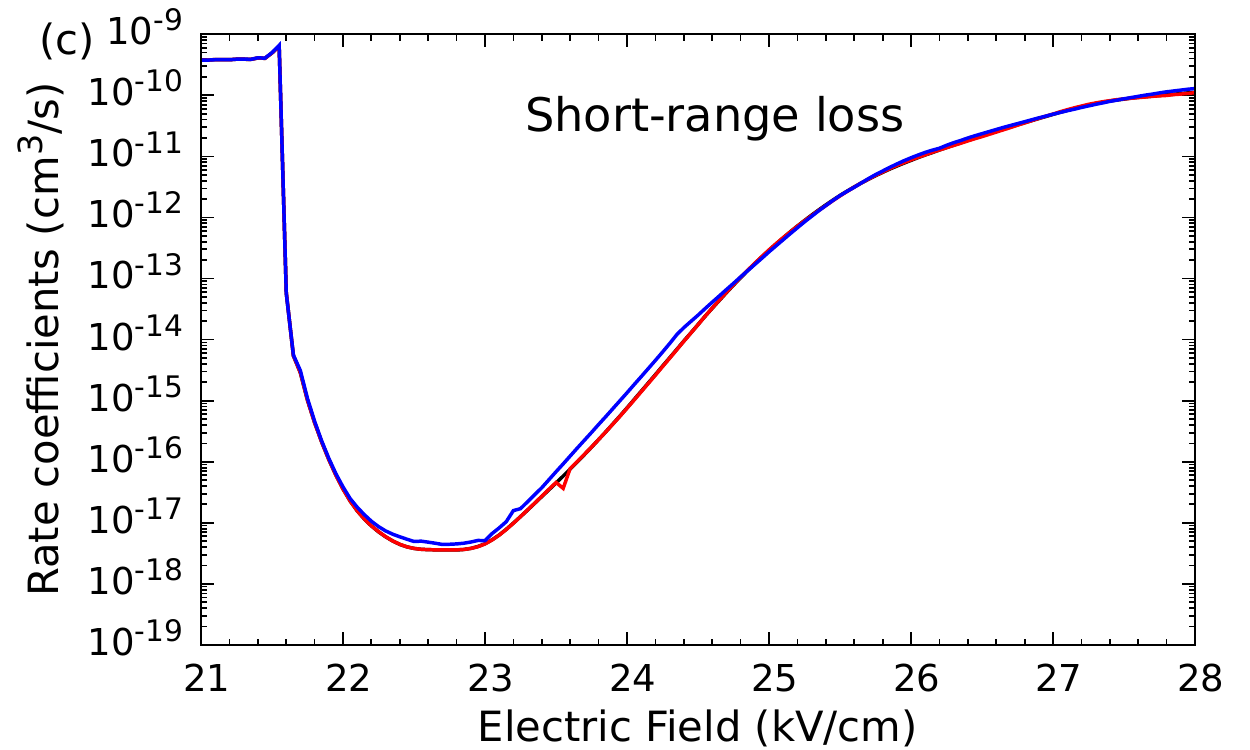}
	}
%    \vspace{-0.8 cm}
	\subfloat[]{
		\includegraphics[width=0.45\textwidth]{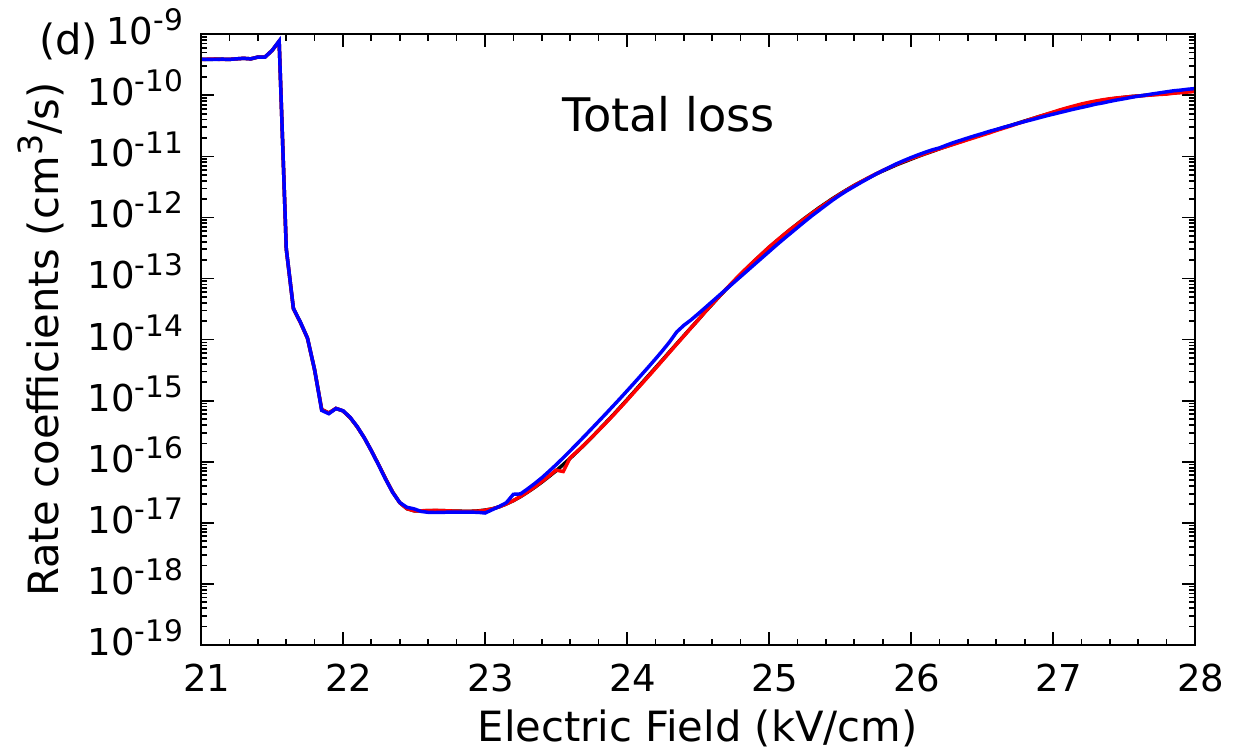}
	}
%    \vspace{-0.4 cm}
\caption{Rate coefficients as a function of electric field for $R_\textrm{absorb}=10\ a_0$ (black), $50\ a_0$ (red) and $80\ a_0$ (blue). The calculations use the large rotor basis set with $L_\textrm{max}=20$. These results are obtained using $M_\textrm{tot}=0$ at $E_\textrm{coll}/k_{\rm B}=10~\mu$K.
}
\label{fig:Rmin}
\end{figure*}

Figure \ref{fig:Rmin} shows rate coefficients calculated with $R_\textrm{absorb}=10$, 50, and 80~$a_0$. It may be seen that the results for 10 and 50~$a_0$ are almost identical. For $80\ a_0$ some inelastic loss is transferred into short-range loss, but the total loss is much less affected. This reflects the fact that some inelastic loss does take place inside 80~$a_0$, as expected from the adiabats in Fig.\ \ref{fig:adiabats}. When $R_\textrm{absorb}=80\ a_0$, this loss appears in the calculations as short-range loss rather than inelastic loss.

\bibliographystyle{../long_bib}
\bibliography{../all,CaF_shieldingData}
\end{document}